\documentclass[nofootinbib,floats,aps,groupedaddress,preprint]{revtex4}
\usepackage{amssymb,amsmath,amsbsy}
\usepackage{mathrsfs}
\usepackage{axodraw}
\newenvironment{Eqnarray}%
     {\arraycolsep 0.14em\begin{eqnarray}}{\end{eqnarray}}
\def\beq{\begin{equation}}
\def\eeq{\end{equation}}
\def\beqa{\begin{Eqnarray}}
\def\eeqa{\end{Eqnarray}}
\def\ifmath#1{\relax\ifmmode #1\else $#1$\fi}
\def\lsup#1{^{\lower 4pt\hbox{$\scriptstyle#1$}}}
\def\llsup#1{^{\lower 2pt\hbox{$\scriptstyle#1$}}}
\def\half{\ifmath{{\textstyle{\frac{1}{2}}}}}
\def\quarter{\ifmath{{\textstyle{\frac{1}{4}}}}}
\def\lsim{\mathrel{\raise.3ex\hbox{$<$\kern-.75em\lower1ex\hbox{$\sim$}}}}
\def\gsim{\mathrel{\raise.3ex\hbox{$>$\kern-.75em\lower1ex\hbox{$\sim$}}}}
\def\sect#1{section~\ref{#1}}

\def\fig#1{Fig.~\ref{#1}}
\def\eq#1{eq.~(\ref{#1})}
\def\Ref#1{ref.~\cite{#1}}
\def\Refs#1#2{refs.~\cite{#1} and \cite{#2}}

\def\eqs#1#2{eqs.~(\ref{#1}) and (\ref{#2})}

\def\eqst#1#2{eqs.~(\ref{#1})--(\ref{#2})}
\def\eqthree#1#2#3{eqs.~(\ref{#1}), (\ref{#2}) and (\ref{#3})}

\def\Eqst#1#2{Eqs.~(\ref{#1})--(\ref{#2})}
\def\Eqs#1#2{Eqs.~(\ref{#1}) and (\ref{#2})}
\def\Eqst#1#2{Eqs.~(\ref{#1})--(\ref{#2})}
\def\msusy{M_{\rm SUSY}}
\def\msusyy{M^2_{\rm SUSY}}
\def\phm{\phantom{-}}

\begin{document}
\preprint{
\vbox{\vspace*{2cm}
      \hbox{SCIPP-07/16}
      \hbox{arXiV:0711.2890~[hep-ph]}
      \hbox{November, 2007}
}}
\vspace*{3cm}

\title{Hard supersymmetry-breaking ``wrong-Higgs'' couplings \\ of the MSSM}
\author{Howard E.~Haber}
%\email[]{haber@scipp.ucsc.edu}
\author{John D.~Mason}
%\email[]{jdmason@physics.ucsc.edu}
\affiliation{Santa Cruz Institute for Particle Physics  \\
   University of California, Santa Cruz, CA 95064, U.S.A. \\
\vspace{1cm}}

\begin{abstract}
  
In the minimal supersymmetric extension of the Standard Model
(MSSM), if the two Higgs doublets are lighter than some subset of
the superpartners of the Standard Model particles, then it is
possible to integrate out the heavy states to obtain an effective
broken-supersymmetric low-energy Lagrangian.  This Lagrangian can
contain dimension-four gauge invariant Higgs interactions that
violate supersymmetry (SUSY).  The \textit{wrong-Higgs} Yukawa
couplings generated by one-loop radiative corrections are a well
known example of this phenomenon.  In this paper, we examine gauge
invariant gaugino--higgsino--Higgs boson interactions that violate
supersymmetry.  Such wrong-Higgs gaugino couplings can be generated
in models of gauge-mediated SUSY-breaking in which some of the messenger
fields couple to the MSSM Higgs bosons.  In regions of parameter
space where the messenger scale is low and $\tan{\beta}$ is large,
these hard SUSY-breaking operators yield $\tan{\beta}$-enhanced
corrections to tree-level supersymmetric relations in the chargino
and neutralino sectors that can be as large as $56\%$.
We demonstrate how physical observables in the chargino sector can
be used to isolate the $\tan\beta$-enhanced effects derived from the
wrong-Higgs gaugino operators.

\end{abstract}


\maketitle

\section{Introduction}

The minimal supersymmetric Standard Model
(MSSM)~\cite{Nilles:1983ge,Haber:1984rc,Martin:1997ns,dreesroy,baertata}
is a supersymmetric extension of the Standard Model, augmented by the
most general set of dimension-two and dimension-three supersymmetry
(SUSY)-breaking operators (allowed by the rules of
\cite{Girardello:1981wz}).  Without additional assumptions, the
resulting MSSM is governed by 124 independent
parameters~\cite{Dimopoulos:1995ju,Haber:1997if}.  These parameters
are considered placeholders for the unknown (and simpler) fundamental
mechanism of spontaneous SUSY-breaking. Since there are many different
models of fundamental SUSY-breaking~\cite{Luty:2005sn}, 
determining the relations among
these soft parameters is an important step towards determining the
organizing principle governing the fundamental mechanism for
SUSY-breaking.  It is
unlikely that all 124 parameters can ever be measured in future
experiments.  Furthermore, additional parameters enter that depend on
the mechanism that communicates the fundamental SUSY-breaking to the
visible sector of MSSM fields; this often involves hidden sector
physics out of the reach of direct detection by colliders because the
scale of physics that governs the communication of the SUSY-breaking
from the hidden sector to the MSSM is of $\mathcal{O}(100$~TeV)
or greater.  One way to infer information about the soft-SUSY-breaking
Lagrangian and the mediation of SUSY-breaking to the MSSM is through
the measurement of radiative corrections to supersymmetric relations
that are imprinted on the parameters of the theory.

Radiative corrections to supersymmetric relations have been the
subject of many studies.  It is quite useful to consider
the case in which there is a separation of the effective low-energy
SUSY-breaking scale ($\msusy$) and the scale of electroweak
symmetry-breaking (characterized by the Higgs vacuum expectation value
$v=246$~GeV).  In this case, one can construct an 
effective Lagrangian~\cite{Georgi:1994qn,Rothstein:2003mp}
below the scale of SUSY-breaking in which the effects
of the SUSY-breaking one-loop effects appear as corrections to
tree-level relations.  For example, one can consider decoupling 
all superpartners, which results in an
effective low-energy two Higgs doublet model 
(2HDM)~\cite{Carena:2002es}
below the SUSY-breaking scale. 
Although the superpartners do not appear in the effective low-energy 2HDM,
their radiative
effects do not decouple and yield predictions of modified relations
between the tree-level Yukawa couplings and the corresponding quark
masses.  These effects can be understood as deriving from the radiatively
corrected MSSM Higgs-Yukawa couplings and the effects of 
radiatively generated so-called \textit{wrong-Higgs} 
Yukawa couplings that violate supersymmetry.

Alternatively, one can consider decoupling only a subset of the
superpartner spectrum and looking at the non-decoupling effects in
both tree-level 2HDM couplings and in the tree-level relations among
the light superpartner couplings.  Such a scenario arises in models of
split-supersymmetry~\cite{Arkani-Hamed:2004fb}. In these models,
the properties of the squarks can be inferred
from deviations of supersymmetric relations between the gauge couplings
and the couplings of the light higgsinos and gauginos
\cite{Giudice:2004tc}.  Although the separation of scales 
between decoupled and non-decoupled states
is essential for the existence of an effective low-energy local
Lagrangian description, such a separation is not required for probes
of SUSY-breaking via radiative effects.  For example, the
deviation of the supersymmetric relations between the gauge and
gaugino couplings were analyzed in~\cite{Katz:1998br,Kiyoura:1998yt}
even though the squarks were not decoupled from the low energy
spectrum.

What is remarkable about the examples cited
above is the role played by dimension-four hard SUSY-breaking
operators.  The coefficients of these operators  
in the effective low-energy Lagrangian below the scale $\msusy$ are 
suppressed by a coupling constant and a loop factor.  In contrast,
the coefficients of dimension-four hard SUSY-breaking operators of
the Lagrangian above $\msusy$ are typically
suppressed by one or two powers of $F/M^2$~\cite{Martin:1999hc}, 
where $F^{1/2}$ characterizes
the fundamental scale of SUSY-breaking and $M$ is the scale of the
physics that transmits SUSY-breaking to the sector of MSSM 
fields.  For example, in cases of
gravity-mediated SUSY-breaking where $F/M\sim \msusy$ and $M$ is the
Planck scale, dimension-four hard SUSY-breaking operators are
Planck-scale suppressed and hence completely negligible.
In models of gauge-mediated supersymmetry 
breaking~\cite{Dine:1981za,Dine:1981gu,Dimopoulos:1981au,Dimopoulos:1982gm,
Nappi:1982hm,AlvarezGaume:1981wy,Dine:1993yw,Dine:1994vc,Dine:1995ag,Giudice:1998bp},
$(g^2/16\pi^2)F/M\sim\msusy$, where $g$ is the relevant gauge coupling
constant and $M$ is a typical mass of the messenger fields that in
some cases can be as low as a few TeV.  In the latter scenario, $F/M^2$
is a rather mild suppression, in which case the corresponding
dimension-four hard SUSY-breaking operators can be phenomenologically
relevant.

This paper is organized as follows.  In \sect{sec:deltab}, we review
the radiative generation of the wrong-Higgs Yukawa couplings of the
effective 2HDM Lagrangian after decoupling the heavy supersymmetric
particles.  The wrong-Higgs Yukawa couplings are
hard SUSY--breaking dimension-four operators that appear in
the effective low-energy theory at the electroweak scale.  One notable
consequence of the wrong-Higgs-couplings is an enhanced correction to
the relation between the bottom quark Yukawa coupling and the bottom
quark mass in the limit of a large ratio of Higgs vacuum expectation
values, $\tan\beta$.  This enhancement can yield a radiative
correction to the Higgs decay rate to bottom-quark pairs that is
significantly larger than the expected size of a one-loop radiative
effect.  Detection of such a deviation would provide insight into the
structure of SUSY-breaking, even while probing interactions
at scales below the heavy masses of the MSSM spectrum.

In \sect{wronghiggsgaugino}, we examine the possibility of analogous
wrong-Higgs interactions that couple gauginos to the higgsinos and
Higgs bosons.  These gaugino--higgsino--Higgs boson interactions are
gauge invariant with respect to the Standard Model gauge group
but are SUSY-breaking, and thus are constrained
to be zero at tree-level in the MSSM. Since we are aiming to use an
effective Lagrangian description of the chargino/neutralino sector at
a scale below the SUSY-breaking scale, we look for regions of MSSM
parameter space where threshold corrections from heavy MSSM particles
can generate these effective operators at one-loop. We show that a
consistent effective Lagrangian treatment of these operators cannot be
achieved from decoupling any subset of MSSM fields at some high 
SUSY-breaking scale.  Nevertheless, when we
parameterize a simple low-energy gauge mediated messenger sector with
couplings to the Higgs doublets, integrating out the messengers does
generate the SUSY-breaking wrong-Higgs operators of interest. 
Models with such messenger interactions have been suggested 
in~\cite{Dvali:1996cu}. For our
purposes, we note that the quantum numbers of the
messenger fields typically
allow for supersymmetric and gauge invariant interactions
with the Higgs doublets.  In this paper, we have explored in detail 
some of the detectable non-decoupling effects of such interactions.

After the messenger sector is integrated out and new gaugino couplings
are present in the effective Lagrangian, corrections to
the off-diagonal elements of the chargino and neutralino mass
matrices are generated.  
In \sect{measure}, we focus on the impact of the wrong-Higgs
gaugino operators on the chargino mass matrix. These SUSY-breaking
interactions will result in deviations in
the tree-level supersymmetric relations between the off diagonal
elements of the chargino mass matrix, the $W$-mass, and $\tan{\beta}$.
We identify one particular correction that is
$\tan{\beta}$-enhanced and dominates over all other
one-loop corrections.  We briefly indicate how the effects of 
the $\tan\beta$-enhanced correction can be isolated in precision
chargino studies at future collider.
Finally, in \sect{nonlocal}, we demonstrate that 
the  $\tan\beta$-enhanced effects of the local wrong-Higgs operators
are parametrically larger than any non-local effects
that could in principle wash-out such effects.
Conclusions and future directions of this work are outlined in the
final \sect{conclude}.

\section{Wrong-Higgs Interactions and the Bottom Quark Mass}
\label{sec:deltab}

The tree-level MSSM Lagrangian consists of SUSY-conserving
mass and interaction terms, supplemented by
soft-SUSY-breaking operators.  Following the rules of
\Ref{Girardello:1981wz}, the soft-SUSY-breaking operators  
include arbitrary dimension-two mass terms and
holomorphic cubic scalar interactions, consistent with the gauge
symmetry of the model.\footnote{Supersymmetry-breaking
mass terms for the fermionic superpartners of scalar fields and
non-holomorphic trilinear scalar interactions 
can potentially destabilize the gauge 
hierarchy~\cite{Girardello:1981wz} in models 
with a gauge-singlet superfield.  
The latter is not present in the MSSM; hence
as noted in \cite{Hall:1990ac,Jack:1999ud}, these so-called non-standard
soft-supersymmetry-breaking terms are benign.
However, the coefficients of these terms (which have dimensions of
mass) are expected to be significantly suppressed compared to the TeV-scale
in a fundamental theory of high-scale supersymmetry-breaking.}  
In particular, all tree-level dimension-four
gauge invariant interactions must respect supersymmetry.

When supersymmetry is broken, in principle all SUSY-breaking
operators consistent with gauge invariance can be generated in the
effective low-energy theory (below the scale of
SUSY-breaking).  The MSSM Higgs sector provides an especially
illuminating example of this phenomenon.  The MSSM contains two
complex Higgs doublet fields $H_u$ and $H_d$ of hypercharge $\pm 1$,
respectively.  The tree-level Higgs--quark
Yukawa Lagrangian is given by:
\beq
\mathscr{L}_{\rm yuk}^{\rm tree}=
-\epsilon_{ij}h_b H_d^i\psi_Q^j\psi_D
+\epsilon_{ij}h_t H_u^i\psi_Q^j\psi_U+{\rm h.c.}\,,
\eeq
where we use two-component notation for the quark
fields.\footnote{Under SU(3)$\times$SU(2)$\times$U(1), the
quantum numbers of the two-component quark fields and Higgs fields are
given by: $\psi_Q(\boldsymbol{3},\boldsymbol{2},1/3)$,
$\psi_U(\boldsymbol{3^*},\boldsymbol{1},-4/3)$,
$\psi_D(\boldsymbol{3^*},\boldsymbol{1},2/3)$,
$H_d(\boldsymbol{1},\boldsymbol{2},-1)$ and
$H_u(\boldsymbol{1},\boldsymbol{2},1)$, where the
electric charge $Q$ (in units of $e$) of the fields are
related to the corresponding isospin $T_3$ and
U(1)-hypercharge ($Y$) by $Q=T_3+\half Y$.  The two-component spinor
product is defined by $\psi\chi\equiv \psi^\alpha\chi_\alpha
=\epsilon^{\alpha\beta}\psi_\beta\chi_\alpha$ ($\alpha$, $\beta=1,2$)
and $\epsilon^{\alpha\beta}$ is antisymmetric with $\epsilon^{12}=1$.
The antisymmetric tensor $\epsilon_{ij}$ (with $\epsilon_{12}=1$)
contracts the gauge SU(2) indices.  We denote the Yukawa couplings
by $h_b$ and $h_t$ (instead of $h_U$ and $h_D$) to emphasize that
the third generation Yukawa couplings dominate those of the lighter two
generations.}
Note that the supersymmetry
restricts the form of the tree-level Yukawa Lagrangian to the
so-called Type-II Yukawa interactions~\cite{Hall:1981bc,hhg}
of the two-Higgs doublet model (2HDM),
in which the neutral component of $H_d$ [$H_u$] couples exclusively to
down-type [up-type] quarks.  Two other possible dimension-four
gauge-invariant non-holomorphic Higgs-quark
interactions terms,  the so-called \textit{wrong-Higgs interactions}
$H_u^{k*} \psi_{d} \psi_{Q}^k$  and $H_d^{k*}\psi_u\psi_Q^k$,
are not supersymmetric (since the dimension-four supersymmetric Yukawa
interactions must be holomorphic), and hence
are absent from the tree-level Yukawa Lagrangian.

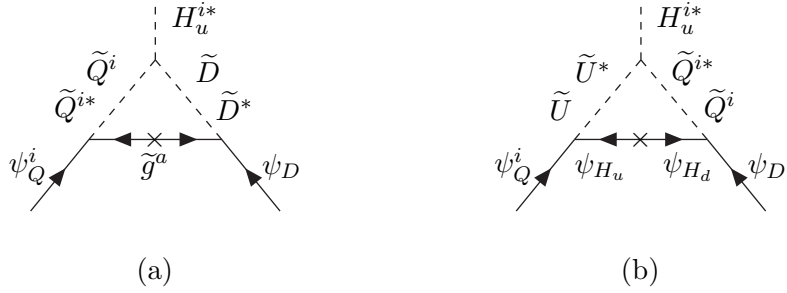
\begin{figure}[t]
\begin{center}
\begin{picture}(100,100)(40,0)
\DashLine(50,80)(50,60){3}
\ArrowLine(3,3)(25,30)
\ArrowLine(97,3)(75,30)
\ArrowLine(50,30)(75,30)
\ArrowLine(50,30)(25,30)
\DashLine(75,30)(50,60){3}
\DashLine(25,30)(50,60){3}
\Text(65,76)[]{$H_u^{i*}$}
\Text(20,43)[]{$\widetilde{Q}^{i*}$}
\Text(30,57)[]{$\widetilde{Q}^{i}$}
\Text(80,43)[]{$\widetilde{D}^{*}$}
\Text(70,57)[]{$\widetilde{D}$}
\Text(50,30)[]{$\times$}
\Text(50,20)[]{$\widetilde g^a$}
\Text(2,20)[]{$\psi_{Q}^i$}
\Text(98,20)[]{$\psi_D$}
\Text(50,-20)[]{(a)}
\end{picture}
\begin{picture}(100,100)(-40,0)
\DashLine(50,80)(50,60){3}
\ArrowLine(3,3)(25,30)
\ArrowLine(97,3)(75,30)
\ArrowLine(50,30)(75,30)
\ArrowLine(50,30)(25,30)
\DashLine(75,30)(50,60){3}
\DashLine(25,30)(50,60){3}
\Text(65,76)[]{$H_u^{i*}$}
\Text(20,43)[]{$\widetilde{U}$}
\Text(32,57)[]{$\widetilde{U}^*$}
\Text(80,43)[]{$\widetilde{Q}^{i}$}
\Text(70,57)[]{$\widetilde{Q}^{i*}$}
\Text(50,30)[]{$\times$}
\Text(35,20)[]{$\psi_{H_u}$}
\Text(68,20)[]{$\psi_{H_d}$}
\Text(2,20)[]{$\psi_{Q}^i$}
\Text(98,20)[]{$\psi_D$}
\Text(50,-20)[]{(b)}
\end{picture}
\end{center}
\caption{\label{wronghiggs}  One-loop diagrams contributing to the
wrong-Higgs Yukawa effective operators.  In (a), the cross ($\times$)
corresponds to a factor of the gluino mass $M_3$.  In (b), the
cross corresponds to a factor of the higgsino Majorana mass
parameter $\mu$. Field labels correspond to annihilation of
the corresponding particle at each vertex of the triangle.}
\end{figure}

Nevertheless, the wrong-Higgs interactions can be generated
in the effective low-energy theory below the scale of
SUSY-breaking.  In particular, one-loop radiative
corrections, in which supersymmetric particles (squarks,
higgsinos and gauginos) propagate inside the 
loop~\cite{Dabelstein:1995js,Coarasa:1995yg,Jimenez:1995wf,Bartl:1995tx,Pierce:1996zz,Borzumati:1999sp,Eberl:1999he,Heinemeyer:2000fa,Haber:2000kq}, 
can generate the
wrong-Higgs interactions as shown in \fig{wronghiggs}.  In constructing 
the one-loop diagrams that produce the wrong-Higgs interactions,
the relevant vertices derive from the following terms of the MSSM Lagrangian.
First, we have the three-scalar interactions:
\beq \label{muanda}
\mathscr{L}_{\rm int}= \mu h_t H_d^i \widetilde Q^{i\,*}\widetilde U^*
+\mu h_b H_u^i\widetilde Q^{i\,*}\widetilde D^*-\epsilon_{ij}\left[
h_b A_b H_d^i\widetilde Q^j\widetilde D-h_t A_t H_u^i\widetilde
Q^j\widetilde U
\right]+{\rm h.c.}\,,
\eeq
which derive from the $\mu$-term of the superpotential and
the soft-SUSY-breaking trilinear scalar interactions (the so-called $A$-terms).
Second, we have the gaugino-quark-squark interactions:
\beqa \label{kahlerint}
\!\!\!\!\!\!\mathscr{L}_{\rm int}&=&
-i\sqrt{2}g_s(\overline{\widetilde g}\lsup{a}\bar\psi_{Q\,k}^i T^a_{k\ell}
\widetilde Q^i_\ell+\widetilde g^a\psi_{U\,k} T^a_{k\ell}\widetilde U^*_\ell
+\widetilde g^a\psi_{D\,k} T^a_{k\ell}\widetilde D^*_\ell+{\rm h.c.})
\nonumber \\
&&-i\sqrt{2}g(\bar\lambda^a\bar\psi_{Q}^i \half\tau^a_{ij}
\widetilde Q^j+{\rm h.c.})
 -i\sqrt{2}g^{\,\prime}\left[y_Q\bar\lambda'\bar\psi_Q^i\widetilde Q^i
+y_U\bar\lambda'\bar\psi_U\widetilde U+y_D\bar\lambda'\bar\psi_D\widetilde D
+{\rm h.c.}\right]\,,
\eeqa %
which derive from the K\"ahler term [cf.~\eq{kahler}].  In
\eq{kahlerint}, $g_s$, $g$ and $g^{\,\prime}$ are the
SU(3)$\times$SU(2)$\times$U(1)$_{\rm Y}$ gauge couplings, $k$ and
$\ell$ are SU(3) color indices and $T^a$ are the SU(3) generators, $i$
and $j$ are the SU(2) gauge indices and $\tau^a$ are the Pauli
matrices, and $y_Q=1/3$, $y_U=-4/3$ and $y_D=2/3$ are the
corresponding hypercharges.  Finally, the higgsino-quark-squark
interactions are the supersymmetric analogs of the Higgs-quark Yukawa
couplings:
\beq \label{susyyuk}
\mathscr{L}_{\rm int}=\epsilon_{ij}\left[h_b \psi_{H_d}^i(\psi_Q^j\widetilde D
+\psi_D\widetilde Q^j)-h_t \psi_{H_u}^i(\psi_Q^j\widetilde U
+\psi_U\widetilde Q^j)+{\rm h.c.}\right]\,.
\eeq

If the squarks are heavy, then one can derive an effective
field theory description of the Higgs-quark Yukawa couplings
below the scale of the heavy squarks, where one has
integrated out the heavy squarks propagating in the loops.
The resulting effective Lagrangian is~\cite{Carena:1999py,Carena:2002es}:
\begin{Eqnarray} \label{yukefflag}
 \mathscr{L}_{\rm yuk}^{\rm eff} = -\epsilon_{ij}(h_b+ \delta
 h_b)\psi_{b}H_d^i
\psi_Q^j  + \Delta h_b \psi_{b} H_u^{k*} \psi_{Q}^k
+ \epsilon_{ij}(h_t + \delta h_t)\psi_{t}H_u^i\psi_Q^j
+ \Delta h_t \psi_tH_d^{k*}\psi_Q^k\,.
 \end{Eqnarray}%
Note that in addition to 
$\delta h_t$ and $\delta h_b$
(which renormalize the Type-II Higgs-quark Yukawa
interactions), wrong-Higgs Yukawa interactions, with coefficients
denoted by $\Delta h_b$ and $\Delta h_t$, have been generated by
the finite loop corrections depicted in \fig{wronghiggs}.
Explicitly, in the limit where the squarks are significantly heavier
than the electroweak symmetry-breaking 
scale~\cite{Hempfling:1993kv,Hall:1993gn,Carena:1994bv,Pierce:1996zz,Carena:1999py},\footnote{We neglect
the contribution of the SU(2)$\times$U(1) gauginos to the
one-loop graphs of \fig{wronghiggs} as these effects are subdominant
to the gluino contribution.}
\begin{Eqnarray} \label{deltahb}
 \Delta h _b = h_b\left[\frac{2\alpha_s}{3\pi}\mu M_3
 \mathscr{I}(M_{\tilde b_1},M_{\tilde b_2},
 M_g) + \frac{h_t}{16\pi^2}\mu A_t \mathscr{I}
(M_{\tilde t_1}, M_{\tilde t_2}, \mu)\right]\,,
 \end{Eqnarray}%
 and
\begin{Eqnarray} \label{iabc}
 \mathscr{I}(a,b,c) = \frac{a^2b^2\ln{(a^2/b^2)} +
   b^2c^2\ln{(b^2/c^2)}
+ c^2a^2\ln{(c^2/a^2)}}{(a^2-b^2)(b^2-c^2)(a^2-c^2)} \,.
\end{Eqnarray}%
In \eq{deltahb}, $M_3$ is the Majorana gluino mass,
$\mu$ is the supersymmetric Higgs-mass parameter, and $\widetilde b_{1,2}$
and $\widetilde t_{1,2}$ are the mass-eigenstate bottom squarks and top
squarks, respectively (our notation follows that of ~\Ref{susypdg}).
Note that $\mathscr{I}(a,b,c)\sim 1/{\rm max}(a^2,b^2,c^2)$
in the limit where at least one of the arguments of
$\mathscr{I}(a,b,c)$ is large.  If $a=b=c$, then
$\mathscr{I}(a,a,a)=1/(2a^2)$.

As expected, the coefficients of the
non-holomorphic dimension-four operators in \eq{yukefflag} vanish in the
supersymmetric limit
(i.e., when the SUSY-breaking parameters $A_b$, $A_t$, and
$M_3$ vanish). Moreover, it is useful
to keep track of the $U(1)_R$-charges of the various operators
appearing in the effective Lagrangian~\cite{jterning}.  All supersymmetric
terms must have total R-charge equal to zero.\footnote{Note that the
dimension-four terms of the tree-level
Lagrangian of a spontaneously-broken supersymmetric model respect the
supersymmetry, and consequently these terms must have zero R charge.}
If we assign the R-charges of the Higgs and quark fields such that
$R(H_u)=R(H_d)=1$ and $R(\psi_Q)=R(\psi_{U})=R(\psi_{D})
=-\frac{1}{2}$ then,
all dimension-four Yukawa interactions
of the tree-level Lagrangian have R-charge zero.
In contrast, the wrong-Higgs Yukawa interactions are operators with
R-charge 2.

We now demonstrate that the effect of the wrong-Higgs couplings is
a $\tan\beta$-enhanced modification of a physical observable.  The
Higgs fields in \eq{yukefflag} can be re-written in terms of the
physical mass-eigenstate neutral and charged Higgs fields and the
Goldstone boson fields~\cite{Gunion:1984yn,Haber:1991mc}:
\begin{Eqnarray}
H_d^1 &=&
\frac{1}{\sqrt{2}}(v\cos\beta+H^0\cos{\alpha} - h^0\sin{\alpha} +
iA^0\sin{\beta}-iG^0\cos\beta)\,,
\\[6pt]
H_u^2 &=&
\frac{1}{\sqrt{2}}(v\sin\beta+H^0\sin{\alpha} + h^0\cos{\alpha} +
iA^0\cos{\beta}+iG^0\sin\beta)\,, \\[6pt]
H_d^2 &=&  H^-\sin{\beta}-G^-\cos\beta\,,\\[6pt]
H_u^1 &=& H^+ \cos{\beta}+G^+\sin\beta \,,
\end{Eqnarray}%
where $v^2\equiv v_u^2+v_d^2=(246~{\rm GeV})^2$ and $\tan\beta\equiv
v_u/v_d$.  Inserting these expressions into \eq{wronghiggs},
we can identify the bottom quark mass as:
\begin{Eqnarray}
m_b = \frac{h_bv}{\sqrt{2}} \cos \beta \left(1 + \frac{\delta h_b}{h_b}
+ \frac{\Delta h_b \tan \beta}{h_b}\right)
\equiv \frac{h_bv}{\sqrt{2}}\cos \beta (1 + \Delta_b)\,,
\end{Eqnarray}%
which defines the quantity $\Delta_b$.  Note that the correction
$\Delta_b$ is $\tan\beta$--enhanced if $\tan\beta\gg 1$. Typically
in the limit of large $\tan\beta$ the term proportional to $\delta
h_b$ can be neglected,
in which case, $\Delta_b\simeq (\Delta h_b/h_b)\tan\beta$.

It is especially noteworthy that the contributions of heavy
supersymmetric particles propagating in the loops of \fig{wronghiggs}
do \textit{not} decouple in the limit of very heavy supersymmetric
particle masses when the quantities $\mu M_3/M_{\tilde q}^2$ and $\mu
A_t/M_{\tilde q}^2$ ($q=t$ or $b$) that appear in \eq{deltahb} are of
$\mathcal{O}(1)$.  Thus, $\Delta_b$ can in principle provide
information about the heavy supersymmetric sector even if the
supersymmetric particles are too heavy to be directly produced at the
LHC.  As $\Delta_b$ is $\tan\beta$-enhanced, one has the possibility
of extracting this quantity from data by measuring the values of the
bottom quark Yukawa coupling, the bottom mass, and $\tan{\beta}$ 
at future colliders in a
precision Higgs program~\cite{Carena:2001bg}.

We now investigate whether it is possible to implement a similar
strategy of probing the heavy sector of supersymmetric models in
studies of the gaugino sector.

\section{Wrong-Higgs interactions in the gaugino sector}
\label{wronghiggsgaugino}

In the MSSM, the supersymmetric partners of the gauge interactions of
charged matter fields (either scalars or fermions) are dimension-four
interactions that couple gauginos to fermions and the scalar
superpartners (the sfermions).  As in the case of the Yukawa
Higgs-fermion interactions, only a subset of all possible
dimension-four gauge invariant gaugino--fermion--sfermion
interactions are supersymmetric.  Thus, we address the following
question: in the low-energy effective theory below the scale that
characterizes SUSY-breaking, are non-supersymmetric
dimension-four gauge invariant gaugino--fermion--sfermion
interactions generated with appreciable coefficients that can
be probed by precision measurements of low-energy observables?

\subsection{SUSY-violating dimension-four gauge invariant
gaugino--higgsino--Higgs boson interactions}

In a supersymmetric field theory, the tree-level supersymmetric
gaugino--fermion--sfermion interactions originate from the
K\"ahler term~\cite{sweinberg, binetruy}:
\beq \label{kahler}
\mathscr{L}_{K} =  \int d^4\theta\,
\Phi_i^{\dagger}(e^{2gV})_{ij}\Phi_j\, \ni\,
-i\sqrt{2}g_a(\bar\lambda^a\bar\psi_i T^a_{ij}A_j-A_i^*T^a_{ij}\psi_j\lambda^a)
\,,
\eeq
where the $\Phi_i$ are chiral superfields (with physical scalar and
two-component fermion components $A_i$ and $\psi_i$) and $V$ is the
gauge vector superfield (with gaugino component $\lambda$).  We denote
the gauge group generators by $T^a$ and allow for a product group
structure for the gauge group by labeling the gauge coupling with the
index $a$ such that $g_a$ is constant within each simple or U(1)
factor of the full gauge group.

The tree-level MSSM chargino and neutralino mass matrices 
derive from three sources:
(1) a supersymmetric higgsino Majorana mass term that
is proportional to the $\mu$ term,
\begin{Eqnarray}
\mathscr{L}_{\mu} = \mu\int d^2\theta \epsilon_{ij}
\widehat H^i_u \widehat H^j_d + {\rm h.c.},
\end{Eqnarray}%
where $\widehat H_u$ and $\widehat H_d$ are the Higgs superfields
whose scalar and fermionic components are $(H_u\,,\,\psi_{H_u})$
and $(H_d\,,\,\psi_{H_d})$, respectively; (2)
soft SUSY-breaking Majorana gaugino masses:
\beq
\mathscr{L}_{soft} = -M\lambda^a\lambda^a - M'\lambda'\lambda' + {\rm h.c.}\,,
\eeq
and (3) the gaugino--higgsino--Higgs boson interactions that arise
from \eq{kahler}.  Contributions to the chargino and neutralino masses
are generated from the latter when the neutral Higgs fields acquire
vacuum expectation values.

Summarizing, after including soft-SUSY-breaking terms, the
gaugino--higgsino--Higgs boson sector of the MSSM Lagrangian (including
mass terms) is given by
\begin{Eqnarray}\label{gMSSM}
\mathscr{L}^{gaugino} & = &
\frac{ig_u}{\sqrt{2}}\lambda^a \tau^a_{ij}\psi^j_{H_u}H_u^{*i}
+  \frac{ig_d}{\sqrt{2}}\lambda^a \tau^a_{ij}\psi_{H_d}^jH_d^{*i}
+  \frac{ig'_u}{\sqrt{2}}\lambda'\psi_{H_u}^iH_u^{*i}
- \frac{ig'_d}{\sqrt{2}}\lambda'\psi_{H_d}^iH_d^{*i} \nonumber \\[6pt]
&&\quad - M\lambda^a\lambda^a - M'\lambda'\lambda'
-  \mu \epsilon_{ij}\psi^i_{H_u}\psi^j_{H_d} + {\rm h.c.}
\end{Eqnarray}%
where
\beq \label{g}
g_u = g_d = g\,,\qquad\qquad g'_u = g'_d = g'\,.
\eeq

Following the strategy of \sect{sec:deltab}, we catalog all possible
dimension-four gauge-invariant operators in the
gaugino--higgsino--Higgs boson sector that violate supersymmetry.
One class of operators of this type are given by:
\beq \label{righthiggs}
\frac{ig_u}{\sqrt{2}}\lambda^a \tau^a_{ij}\psi^j_{H_u}H_u^{*i}
+  \frac{ig_d}{\sqrt{2}}\lambda^a \tau^a_{ij}\psi_{H_d}^jH_d^{*i}
+  \frac{ig'_u}{\sqrt{2}}\lambda'\psi_{H_u}^iH_u^{*i}
- \frac{ig'_d}{\sqrt{2}}\lambda'\psi_{H_d}^iH_d^{*i}+{\rm h.c.}\,,
\eeq
where the coupling $g_u$, $g_d$, $g'_u$ and $g'_d$ deviate from their
supersymmetric values given in \eq{g}.  Such effects are generated
in the one-loop corrections to these interactions.  They have been studied
in detail in \Refs{Katz:1998br}{Kiyoura:1998yt}.  In this paper, we focus
on the following gauge invariant four-dimensional operators that
are not present in the supersymmetric Lagrangian:
\begin{Eqnarray}
&& ik_1\lambda^a\tau^a_{ij}\psi^j_{H_u}\epsilon_{ki}H_d^k\,,\label{k1}\\
&& ik_2 \lambda' \psi_{H_u}^k\epsilon_{ki} H_d^i\,,\label{k2}\\
&& ik_3\lambda^a\tau^a_{ij}\psi_{H_d}^j\epsilon_{ki}H_u^k\,,\label{k3}\\
&& ik_4\lambda'\psi^i_{H_d}\epsilon_{ki}H_u^k\,.\label{k4}
\end{Eqnarray}%
It is straightforward to verify that these are SUSY-breaking operators.
For example, if we assign R-charges to the Higgs superfields so that
$R(\widehat H_u)=R(\widehat H_d)=1$
as before, then the component Higgs fields possess the same R-charges
as their superfield parents, whereas
corresponding higgsino fields have
R-charges $R(\psi_{H_u})=R(\psi_{H_d})=0$.
The vector superfield~$V$ has R-charge equal to zero, which implies that
R-charges of the gaugino fields are given by $R(\lambda)=R(\lambda')=1$.
Consequently, the operators in \eq{righthiggs} all have total R-charge
equal to zero, whereas the operators listed in \eqst{k1}{k4} have
R-charge equal to 2.  Hence, these hard-breaking operators do not
appear in the tree-level MSSM.  Nevertheless, these operators could be
generated radiatively by the threshold
effects of integrating out heavy fields just as the wrong-Higgs
Yukawa couplings to the quarks were generated after integrating out the
superpartners.  We now investigate whether these operators are generated
in the low-energy effective theory at energies below the scale of
SUSY-breaking.

\subsection{Generating wrong-Higgs gaugino operators from a partially
decoupled MSSM}
\label{within}

In the case of the radiative corrections to the bottom quark-Higgs Yukawa
interactions, the effective Lagrangian description was successful
because the one-loop Feynman
graphs with heavy supersymmetric particles propagating in the loops
yielded effective local operators after integrating out the heavy states.
Due to SUSY-breaking effects that generate large mass splitting between
particles and their superpartners, the resulting dimension-four local
operators that survive in the effective low energy theory can violate
supersymmetry; hence the origin of the wrong-Higgs Yukawa couplings.
In the case of gaugino interactions, one cannot usefully
integrate out all the superpartners (in the limit where all superpartners
are heavy), as this would remove the gaugino interaction terms of interest
from the effective low-energy theory. Instead, one must consider
a different limit where a subset of superpartners (\textit{not} including the
Higgs doublets, the gauginos, and the higgsinos) are integrated out.
In this limit, we take $\mu$,
$M$, and $M'$ in \eq{gMSSM} small compared to squark and slepton masses.
In particular, we assume that the soft-SUSY-breaking scalar mass parameters
and $A$-terms (that govern the holomorphic trilinear scalar couplings)
are of $\mathcal{O}(\msusy)$, which we shall take to be
significantly larger than the scale of electroweak symmetry breaking.

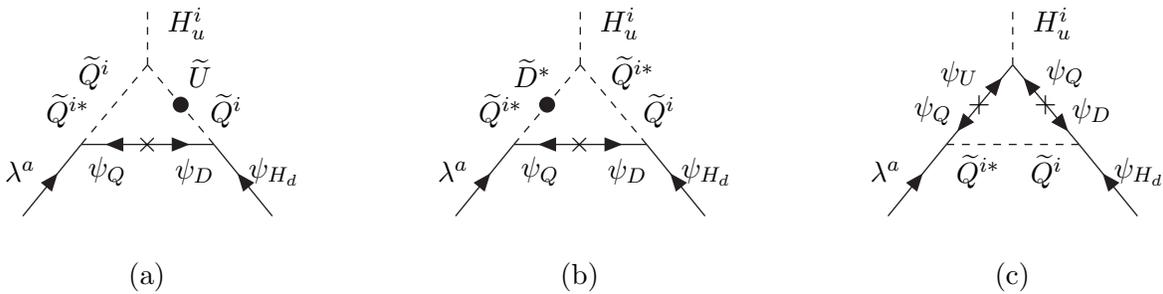
\begin{figure}[t!]
\begin{center}
\begin{picture}(100,100)(60,0)
\DashLine(50,80)(50,60){3}
\ArrowLine(3,3)(25,30)
\ArrowLine(97,3)(75,30)
\ArrowLine(50,30)(75,30)
\ArrowLine(50,30)(25,30)
\DashLine(75,30)(50,60){3}
\DashLine(25,30)(50,60){3}
\Text(65,76)[]{$H_u^{i}$}
\Text(20,43)[]{$\widetilde{Q}^{i*}$}
\Text(30,57)[]{$\widetilde{Q}^{i}$}
\Text(80,43)[]{$\widetilde{Q}^{i}$}
\Text(70,57)[]{$\widetilde{U}$}
\Text(50,30)[]{$\times$}
\Text(35,20)[]{$\psi_Q$}
\Text(68,20)[]{$\psi_D$}
\Text(2,20)[]{$\lambda^a$}
\Text(98,20)[]{$\psi_{H_d}$}
\Vertex(62.5,45){3}
\Text(50,-20)[]{(a)}
\end{picture}
\begin{picture}(100,100)(0,0)
\DashLine(50,80)(50,60){3}
\ArrowLine(3,3)(25,30)
\ArrowLine(97,3)(75,30)
\ArrowLine(50,30)(75,30)
\ArrowLine(50,30)(25,30)
\DashLine(75,30)(50,60){3}
\DashLine(25,30)(50,60){3}
\Text(65,76)[]{$H_u^{i}$}
\Text(20,43)[]{$\widetilde{Q}^{i*}$}
\Text(32,57)[]{$\widetilde{D}^*$}
\Text(80,43)[]{$\widetilde{Q}^{i}$}
\Text(70,57)[]{$\widetilde{Q}^{i*}$}
\Text(50,30)[]{$\times$}
\Text(35,20)[]{$\psi_{Q}$}
\Text(68,20)[]{$\psi_{D}$}
\Text(2,20)[]{$\lambda^a$}
\Text(98,20)[]{$\psi_{H_d}$}
\Vertex(37.5,45){3}
\Text(50,-20)[]{(b)}
\end{picture}
\begin{picture}(100,100)(-60,0)
\DashLine(50,80)(50,60){3}
\ArrowLine(3,3)(25,30)
\ArrowLine(97,3)(75,30)
\DashLine(25,30)(75,30){3}
\ArrowLine(62.5,45)(50,60)
\ArrowLine(62.5,45)(75,30)
\ArrowLine(37.5,45)(50,60)
\ArrowLine(37.5,45)(25,30)
\Text(65,76)[]{$H_u^{i}$}
\Text(20,43)[]{$\psi_Q$}
\Text(30,57)[]{$\psi_U$}
\Text(80,43)[]{$\psi_D$}
\Text(70,57)[]{$\psi_Q$}
\Text(62.5,45)[]{$+$}
\Text(37.5,45)[]{$+$}
\Text(37,20)[]{$\widetilde Q^{i*}$}
\Text(63,20)[]{$\widetilde Q^i$}
\Text(2,20)[]{$\lambda^a$}
\Text(98,20)[]{$\psi_{H_d}$}
\Text(50,-20)[]{(c)}
\end{picture}\end{center}
\caption{\label{failhiggs}  One-loop diagrams contributing to the
wrong-Higgs gaugino effective operators.  The cross ($\times$) indicates
the two-component fermion propagator that is proportional to
the corresponding Dirac mass.  In (a) and (b) the solid dot
indicates an insertion of the Higgs vacuum expectation value.
Field labels correspond to annihilation at each vertex of the triangle.}
\end{figure}

In constructing the one-loop diagrams that produce the wrong-Higgs
gaugino operators, the relevant vertices again derive from the
interaction terms of the MSSM Lagrangian exhibited in
\eqst{muanda}{susyyuk}.  We first attempt to construct graphs
analogous to those of \fig{wronghiggs}.  Two possible graphs that
contribute to the wrong-Higgs gaugino operator that is proportional to
$k_3$ [\eq{k3}] are exhibited in \fig{failhiggs}(a) and (b).  However,
these graphs must contain an insertion of a Higgs vacuum expectation
value at the location of the solid dot on the squark lines
(corresponding to $\widetilde q_L$---$\widetilde q_R$ mixing).  Thus,
simple power counting, under the assumption that the squark masses and
$A$-terms are of order $\msusy\gg v$, implies that the contributions
of \fig{failhiggs}(a) and (b) are of $\mathcal{O}(m_bm_t/\msusyy)$ and
$\mathcal{O}(m^2_b /\msusyy)$, respectively.  Hence, these
contributions decouple in the limit of $\msusy\gg v$.

There is another vertex correction with an internal squark line, shown
in \fig{failhiggs}(c) that can
potentially contribute to the wrong-Higgs gaugino operators.
Simple power counting again
implies that the contribution of
\fig{failhiggs}(c) is of $\mathcal{O}(m_b m_t/\msusyy)$ and hence decouples.
The decoupling properties of \fig{failhiggs} could have been
anticipated due to the insertion of two vacuum expectation values in
each diagram (either via the Dirac mass for the bottom and/or top
quark or the $\widetilde Q$--$\widetilde U$ and/or $\widetilde
Q$--$\widetilde D$ squark mixing).  Hence, replacing the vacuum
expectation value by the appropriate Higgs field, we see that the
contributions of \fig{failhiggs} actually correspond to dimension-six
operators with the expected decoupling behavior.

Similar conclusions also apply to the contributions to the three
other wrong-Higgs gaugino operators  [\eqthree{k1}{k2}{k4}] introduced above.
Consequently, we conclude that there
are no non-decoupling one-loop contributions to the effective
operators in \eqst{k1}{k4} from heavy MSSM fields.

\subsection{Generating wrong-Higgs gaugino operators in a model of
gauge-mediated supersymmetry breaking}
\label{largesquark}

The MSSM is an effective low-energy theory of broken supersymmetry.
One expects that the soft-SUSY-breaking dimension-two and
dimension-three terms of the MSSM Lagrangian are generated by a new
sector of heavy states.  In models of gauge-mediated supersymmetry
breaking (GMSB), super\-symmetry breaking is transmitted to the MSSM
via gauge forces~\cite{Dine:1981za,Dine:1981gu,Dimopoulos:1981au,
Dimopoulos:1982gm,Nappi:1982hm,AlvarezGaume:1981wy,Dine:1993yw,Dine:1994vc,Dine:1995ag,Giudice:1998bp}.  
A typical structure of such models involves a
hidden sector where supersymmetry is broken, a messenger sector
consisting of particles (messengers) with
SU(3)$\times$SU(2)$\times$U(1) quantum numbers, and the visible sector
consisting of the fields of the MSSM.  The
direct coupling of the messengers to the hidden sector generates a
SUSY-breaking spectrum in the messenger sector.  Finally,
super\-symmetry breaking is transmitted to the MSSM via the virtual
exchange of the messenger fields.

In order to maintain the unification of gauge coupling
constants the messengers and the Higgs doublets are taken
to be members of complete irreducible representations of SU(5).
Moreover, for appropriate choices of gauge quantum numbers for
the messenger fields,
it is possible to construct gauge invariant supersymmetric 
direct Yukawa couplings between the Higgs and messenger fields.
Here we consider a model of such interactions and show that
one-loop corrections involving messenger fields in the loop
can generate the wrong-Higgs gaugino operators that survive in
the low-energy theory below the scale of SUSY-breaking.

We begin by parameterizing a simple hidden and messenger sector that
couples to the Higgs doublets.  All the hidden sector
dynamics will be described by a chiral superfield $\widehat Z$ whose scalar
component ($Z$)
and $F$-term component ($F_Z$) acquire vacuum expectation values.
$\widehat Z$~couples to
four messenger superfields, $\widehat M_1$, $\widehat{\overline{M}}_1$,
$\widehat{M}_2$ and $\widehat{\overline{M}}_2$, whose quantum numbers
under SU(3)$\times$SU(2)$\times$U(1)$_{\rm Y}$ are listed in
Table~\ref{tab:mess}.
\begin{table}[t!]
\centering
\begin{tabular}{|c|ccc|} \hline
Superfield & \quad SU(3) \quad &\quad SU(2)\quad & 
\quad U(1)$_{\rm  Y}$ \quad\\ \hline
$\widehat H_d$  & 1 & 2 & $-1$ \\
$\widehat H_u$   & 1 & 2 & $\phm 1$ \\
$\widehat M_1$    & 1 & 2 & $\phm 1$ \\
$\widehat{\overline{M}}_1$  & 1 & 2 & $-1$ \\
$\widehat M_2$  & 1 & 1 & $-2$ \\
$\widehat{\overline{M}}_2$ & 1 & 1 & $\phm 2$ \\ \hline
\end{tabular}
\caption{\label{tab:mess} Gauge quantum numbers of the Higgs and
messenger superfields.}
\end{table}

In GMSB models, it is typically difficult to generate a $\mu$
and $B$ term of the same order of magnitude at the scale
of low-energy SUSY-breaking.  Addressing this
problem is beyond the scope of this paper.  Here we simply note that
messenger loops will generate both $\mu$ and $B$ parameters,
as explained in~\cite{Dvali:1996cu}
and such messenger loops might play a role in a solution to the $\mu$
and $B$ problem as recently discussed in~\cite{Roy:2007nz} and
\cite{Murayama:2007ge}.  Here, we shall focus on
the interactions of the messenger sector and determine the
phenomenological implications of messenger interactions with the
Higgs fields.  A simple superpotential that communicates
SUSY-breaking through gauge mediation with Higgs interactions and is
consistent with the symmetries exhibited in Table~\ref{tab:mess}
is given by:\footnote{Note that this superpotential could 
in principle be embedded in a
grand unified theory, where the various superfields live within
the following SU(5) multiplets:
${\widehat M}_1\subset\boldsymbol{5}$, ${\widehat{\overline{M}}}_1\subset
\boldsymbol{5^*}$, ${\widehat M}_2\subset \boldsymbol{10^*}$,
${\widehat{\overline{M}}}_2 \subset\boldsymbol{10}$, ${\widehat H}_u
\subset\boldsymbol{5}$, and
${\widehat H}_d \subset \boldsymbol{5^*}$.
In this case, the Higgs/messenger couplings would
originate as subsets of the $\boldsymbol{5^* \times 5^* \times
10}$ and $\boldsymbol{5\times 5 \times 10^*}$ couplings.}
\beq \label{spotential}
W= \gamma_1\epsilon_{ij}\widehat Z {\widehat{M}}_1^i
{\widehat{\overline{M}}}_1\lsup{j}
+ \gamma_2 \widehat Z {\widehat{M}}_2{\widehat{\overline{M}}}_2
+\alpha \epsilon_{ij}{\widehat H}_u^i{\widehat M}_1^j{\widehat M}_2 +
\beta \epsilon_{ij}{\widehat H}_d^i{\widehat{\overline{M}}}_1\lsup{j}
{\widehat{\overline{M}}}_2\,.
\eeq
After taking into account the vacuum expectation values of the 
superfield $\widehat Z$, this superpotential
yields masses and interaction terms for the messenger scalar and fermionic
component fields.  For the computations presented in this section, we record
the relevant mass and interaction terms:
\begin{Eqnarray}\label{messmass}
-\mathscr{L}_{\rm mass}&=&
|\gamma_1\langle Z \rangle|^2|M_1|^2 +  |\gamma_1\langle Z
\rangle|^2|\overline{M}_1|^2 +  |\gamma_2\langle Z \rangle|^2|M_2|^2 +
|\gamma_2\langle Z \rangle|^2|\overline{M}_2|^2 \nonumber \\ 
&+&\gamma_1 F_Z \epsilon_{ij}M_1^i\overline{M}_1^j + 
\gamma_2F_Z M_2\overline{M}_2
+\gamma_1\langle Z \rangle\epsilon_{ij}\psi_{M_1}^i\psi_{\bar{M_1}}^j
+ \gamma_2\langle Z \rangle \psi_{M_2}\psi_{\bar{M_2}}+{\rm h.c.}\,,
\end{Eqnarray}%
and
\begin{Eqnarray}\label{messyuk}
\mathscr{L}_{\rm int} &=& -\epsilon_{ij} \biggl[\alpha
\left(H_u^i\psi_{M_1}^j\psi_{M_2}
+ M_1^j\psi_{H_u}^i\psi_{M_2} + M_2\psi_{H_u}^i\psi_{M_1}^j\right)\nonumber \\
&& \qquad +\beta \left(H_d^i\psi_{\bar{M}_1}^j\psi_{\bar{M}_2}
+ \overline{M}_1^j\psi_{H_d}^i\psi_{\bar{M_2}} 
+ \overline{M}_2\psi_{H_d}^i\psi_{\bar{M_1}}^j\right)\biggr]
\nonumber \\[6pt]
&&-\gamma_2\epsilon_{ij}\langle Z \rangle\left[\alpha H_u^i M_1^j
\overline{M}_2\llsup{*}
+\beta  H_d^i\overline{M}_1^j M_2^*\right]
+\gamma_1 \langle Z\rangle \left[\alpha  H_u^i\overline{M}_1\lsup{i*}M_2
-\beta H_d^i M_1^{i*}\overline{M}_2\right]+ {\rm h.c.} \nonumber \\
\phantom{line}
\end{Eqnarray}%
We also record the relevant gaugino--particle--sparticle interactions
involving the messenger scalars and their fermionic superpartners.  
From \eq{kahler}, we obtain:
\beq \label{messgaugino}
\mathscr{L}_{\rm int} = \frac{ig}{\sqrt{2}}\lambda^a\tau^a_{ij}\left[
\psi_{M_1}^jM_1^{i*} +\psi_{\bar{M}_1}^j\overline{M}_1\lsup{i*}\right]
-\frac{ig'}{\sqrt{2}}\lambda'\left[\psi_{M_1}^iM_1^{i*}
- \psi_{\bar{M}_1}^i \overline{M}_1\lsup{i*}
-2\psi_{M_2}M_2^* +2 \psi_{\bar{M}_2}\overline{M}_2\right]+{\rm h.c.}
\eeq

In typical GMSB models, soft-SUSY-breaking masses for
the gauginos are generated at one-loop and soft-SUSY-breaking squared-masses
for the scalars (squarks, sleptons and Higgs bosons) are generated 
at two-loops.  Consequently, the soft-SUSY-breaking masses 
of the gauginos and scalars are of the same order of magnitude.
For example, in order to ensure that $M \sim M' \sim \mu \sim$ 100---500~GeV, 
one must choose $F_Z/{\langle Z \rangle} \sim$ 100~TeV.
However, if the Higgs bosons couple directly to the messengers as in
\eq{spotential}, soft-SUSY-breaking masses for the Higgs fields
and a $B$-term will be generated at \textit{one} loop order.  
In this case, an unnatural fine-tuning is 
required to keep these Higgs soft-masses 
$\lsim\mathcal{O}(\msusy)$. 
In order to reduce the amount of fine-tuning,\footnote{In this
context, we accept the order $1$---$10\%$ fine-tuning
associated with the so-called little hierarchy 
problem~\cite{Cheng:2003ju,Harnik:2003rs,Birkedal:2004xi}.}  
we shall take $F_Z/{\langle Z \rangle} \sim$ 20~TeV.  In
such a model, the contributions of the messenger superfields $\widehat M_1$
and $\widehat M_2$ to slepton and gaugino masses are
phenomenologically too small. One must then add an additional source
of SUSY-breaking to the theory.  An extra pair of 
weak doublet messenger fields (and corresponding color-triplet
messenger fields) coupling to a different spurion $\widehat X$, where 
$\widehat X = \langle X \rangle + \theta^2F_{X}$
and $F_X/{\langle X \rangle} \sim$ 100~TeV is sufficient to raise the
masses of the sleptons, squarks and gauginos above the current experimental 
bounds.  Henceforth, we shall focus exclusively on the radiative
effects of the messenger fields that couple to the spurion $\widehat Z$ 
(and in what follows, the term ``messengers'' will always refer 
to these fields).

Since $\langle Z \rangle$ sets the scale for the average
messenger masses, the consistency of the model\footnote{For
$F_Z\lsim \langle Z \rangle^2$, large splittings of squared-masses
in the messenger sector would drive some scalar squared-masses negative. 
In practice, one requires the masses of all messengers to lie above the
masses of the superpartners of the Standard Model particles.  This
sets an upper bound on $F_Z/{\langle Z \rangle}^2$ of $\mathcal{O}(1)$.}
requires that $F_Z\lsim \langle Z \rangle^2$.  Under our model
assumption, $F_Z/{\langle Z \rangle}\sim 20$~TeV, we can write
$\langle Z \rangle\sim 20~{\rm TeV}/(F_Z/{\langle Z \rangle}^2)$.
There are then two regimes
of possible interest.  If $F_Z\sim \langle Z \rangle^2$, then the
messengers are rather ``light,'' with an average mass 
of order 20~TeV.  In contrast, if $F_Z\ll
\langle Z \rangle^2$, then the messenger masses are significantly
heavier.  

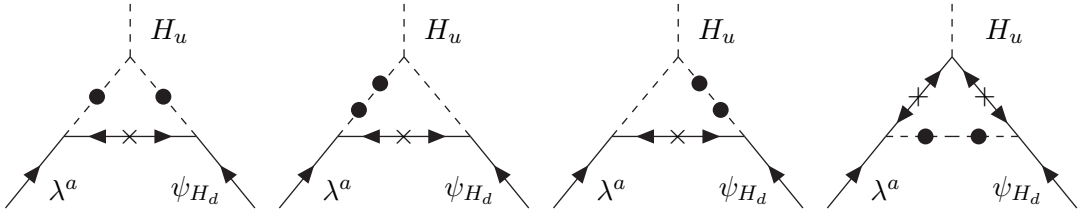
\begin{figure}[t]
\begin{center}
\begin{picture}(100,100)(0,0)
\DashLine(50,80)(50,60){3}
\ArrowLine(3,3)(25,30)
\ArrowLine(97,3)(75,30)
\ArrowLine(50,30)(75,30)
\ArrowLine(50,30)(25,30)
\DashLine(75,30)(50,60){3}
\DashLine(25,30)(50,60){3}
\Vertex(37.5,45){3}
\Vertex(62.5,45){3}
\Text(65,70)[]{$H_u$}
\Text(50,30)[]{$\times$}
\Text(25,10)[]{$\lambda^a$}
\Text(75,10)[]{$\psi_{H_{d}}$}
\end{picture}
\begin{picture}(100,100)(0,0)
\DashLine(50,80)(50,60){3}
\ArrowLine(3,3)(25,30)
\ArrowLine(97,3)(75,30)
\ArrowLine(50,30)(75,30)
\ArrowLine(50,30)(25,30)
\DashLine(75,30)(50,60){3}
\DashLine(25,30)(50,60){3}
\Vertex(33,40){3}
\Vertex(41,50){3}
\Text(65,70)[]{$H_u$}
\Text(50,30)[]{$\times$}
\Text(25,10)[]{$\lambda^a$}
\Text(75,10)[]{$\psi_{H_{d}}$}
\end{picture}
\begin{picture}(100,100)(0,0)
\DashLine(50,80)(50,60){3}
\ArrowLine(3,3)(25,30)
\ArrowLine(97,3)(75,30)
\ArrowLine(50,30)(75,30)
\ArrowLine(50,30)(25,30)
\DashLine(75,30)(50,60){3}
\DashLine(25,30)(50,60){3}
\Vertex(58,50){3}
\Vertex(66,40){3}
\Text(65,70)[]{$H_u$}
\Text(50,30)[]{$\times$}
\Text(25,10)[]{$\lambda^a$}
\Text(75,10)[]{$\psi_{H_{d}}$}
\end{picture}
\begin{picture}(100,100)(0,0)
\DashLine(50,80)(50,60){3}
\ArrowLine(3,3)(25,30)
\ArrowLine(97,3)(75,30)
\DashLine(75,30)(50,30){3}
\DashLine(25,30)(50,30){3}
\ArrowLine(62.5,45)(75,30)
\ArrowLine(62.5,45)(50,60)
\ArrowLine(37.5,45)(25,30)
\ArrowLine(37.5,45)(50,60)
\Text(70,70)[]{$H_u$}
\Text(37.5,45)[]{$+$}
\Text(62.5,45)[]{$+$}
\Text(25,10)[]{$\lambda^a$}
\Text(75,10)[]{$\psi_{H_d}$}
\Vertex(40,30){3}
\Vertex(60,30){3}
\end{picture}
\caption{\label{higgsmess} One-loop diagrams with internal lines consisting of
scalar and fermionic messenger fields.  The cross ($\times$) indicates
the two-component fermion propagator that is proportional to
the corresponding Dirac mass.  The solid dot indicates
an $F_Z$ mass-insertion on the scalar messenger line.}
\end{center}
\end{figure}

Consider first the case of $F_Z\ll \langle Z \rangle^2$.  In this
case, the mass splittings of $M_1$, $\overline{M}_1$, $M_2$, and
$\overline{M}_2$ can be treated as perturbations about the average
mass $\langle Z \rangle$.  Let us examine the contributions to the
SUSY-breaking wrong-Higgs gaugino interactions generated by
integrating out the messenger fields.  In this case we can evaluate
the diagrams of \fig{higgsmess} in the mass-insertion approximation,
with messengers running in the loops and mass insertions of $F_Z$ on
the scalar propagator lines.

The graphs with two scalar propagators enter with the opposite sign
compared to the graph with the two fermion propagators. We find the
following leading contribution to $k_3$ in the mass-insertion approximation:
\beq \label{fzlarge}
k_3 \sim\frac{g}{16\pi^2}
\left( \frac{F_Z}{\langle Z \rangle^2} \right)^2\,,
\eeq
where we have suppressed an overall numerical coefficient of $\mathcal{O}(1)$.
The contribution to $k_3$ (and similarly for the other $k_i$)
decouples in the limit of ${F_Z}\ll{\langle Z \rangle^2}$,
in agreement with the expectations of~\cite{Martin:1999hc}.   
In \sect{measure}, we will exhibit a correction to a physical quantity
that is proportional to $k_3\tan\beta$.  However, even with
a $\tan\beta$-enhancement as large as 50, the ultimate effect of such
corrections is too small to be observed.

If $F_Z \sim \langle Z \rangle^2$, the mass-insertion approximation employed
above is no longer valid.  The results of \eq{fzlarge} are suggestive
of the possibility that the effect of integrating out the messenger
fields could produce non-decoupling contributions to the $k_i$ in the
low-energy effective theory below the messenger scale.  
In order to evaluate the one-loop contributions to $k_i$
for $F_Z \sim \langle Z \rangle^2$, we must employ 
mass-eigenstates for the scalar messengers that appear in the loops
of \fig{higgsmess}.
Scalar messengers must be re-expressed in terms of their mass 
eigenstates, and the
diagrams must be evaluated with internal line mass eigenstates.
In this paper, we present the explicit computation
for $k_3$, as this coefficient is 
is the only one that governs unsuppressed
corrections to the chargino mass matrix and interactions.

From \eq{messmass}, it follows that the
fermionic messenger fields organize themselves into
two Dirac fermions $\Psi_1\equiv(\psi_{M_1}\,,\,\bar\psi_{\bar{M}_1})$
and $\Psi_2\equiv(\psi_{M_2}\,,\,\bar\psi_{\bar{M}_2})$ with 
corresponding Dirac masses $m_1\equiv\gamma_1\langle Z\rangle$ 
and $m_2\equiv\gamma_2\langle Z\rangle$.
Moreover, we can express the mass-eigenstate scalar messengers in terms of 
the corresponding interaction-eigenstate fields.  We work in
the limit of exact SU(2)$\times$U(1) and ignore small 
corrections to the messenger masses of order the electroweak
scale that are generated by the neutral Higgs vacuum expectation values.
It is convenient to rewrite the complex scalar messenger fields
in terms of their real and imaginary parts:
\beq
M_1^i=\frac{1}{\sqrt{2}}(M^i_{1R}+M^i_{1I})\,,\qquad\qquad
M_2=\frac{1}{\sqrt{2}}(M_{2R}+M^i_{2I})\,,
\eeq
and similarly for the barred fields $\overline{M}_1^i$ and
$\overline{M}_2$.  From \eq{messmass}, the scalar messenger
mass-eigenstates are determined from:
\beqa
-\mathscr{L}_{\rm mass}&=&\half\left(M^i_{1R}\quad
\epsilon_{ij}\overline{M}\lsup{\,j}_{1R}\right)\boldsymbol{\Gamma}_+^{(1)}
\begin{pmatrix} M^i_{1R}\\
\epsilon_{ik}\overline{M}\lsup{\,k}_{1R}\end{pmatrix}
+\half\left(M^i_{1I}\quad
\epsilon_{ij}\overline{M}\lsup{\,j}_{1I}\right)\boldsymbol{\Gamma}_-^{(1)}
\begin{pmatrix} M^i_{1I}\\
\epsilon_{ik}\overline{M}\lsup{\,k}_{1I}\end{pmatrix} \nonumber \\[8pt]
&& \quad +\half\left(M_{2R}\quad
\overline{M}_{2R}\right)\boldsymbol{\Gamma}_+^{(2)}
\begin{pmatrix} M_{2R}\\ \overline{M}_{2R}\end{pmatrix}
+\half\left(M_{2I}\quad
\overline{M}_{2I}\right)\boldsymbol{\Gamma}_-^{(2)}
\begin{pmatrix} M_{2I}\\ \overline{M}_{2I}\end{pmatrix}\,,
\eeqa%
where
\beq
\boldsymbol{\Gamma_{\pm}^{(n)}}=\begin{pmatrix} 
\gamma_n^2\langle Z\rangle^2 \quad & \pm\gamma_n F_Z\\
\pm\gamma_n F_Z \quad & \gamma_n^2\langle Z\rangle^2
\end{pmatrix}\,,\qquad\quad (n=1,2)\,.
\eeq
The scalar messenger mass eigenstates are:
\beq
M_{\pm 1{R,I}}^i\equiv \frac{1}{\sqrt{2}}
\left(M_{1R,I}^i\pm \epsilon_{ij}\overline{M}\lsup{j}_{1R,I}\right)\,,
\qquad\quad 
M_{\pm 2{R,I}}\equiv \frac{1}{\sqrt{2}}
\left(M_{2R,I}\pm \overline{M}_{2R,I}\right)\,,
\eeq
with corresponding  masses $m^\pm_{nR,I}$ given by (for $n=1,2$):
\beqa
(m^-_{1I})^2&=&(m^+_{1R})^2=\gamma_1^2\langle Z\rangle+\gamma_1 F_Z\,, 
\qquad\qquad
(m^-_{2I})^2=(m^+_{2R})^2=\gamma_2^2\langle Z\rangle+\gamma_2 F_Z\,,
\\ 
(m^+_{1I})^2&=&(m^-_{1R})^2=\gamma_1^2\langle Z\rangle-\gamma_1 F_Z\,, 
\qquad\qquad
(m^+_{2I})^2=(m^-_{2R})^2=\gamma_2^2\langle Z\rangle-\gamma_2 F_Z\,.
\eeqa% 

We can then evaluate the exact one-loop threshold corrections contributing to
$k_3$ by employing Feynman rules with messenger mass-eigenstates. 
In the limit where the internal particle masses are
much greater than the external momenta,
\beqa \label{k3eff}
\frac{k_3}{g} &  = &  \frac{\alpha\beta (\gamma_2+\gamma_1)}
{128\sqrt{2}\pi^2}m_1
\langle Z \rangle \biggl[ \mathscr{I}(m_1,m^+_{1R},m^+_{2R})
+\mathscr{I}(m_1,m^+_{1I},m^+_{2I}) +
\mathscr{I}(m_1,m^+_{1R},m^-_{2I}) \nonumber \\
&&\qquad\qquad\qquad + \mathscr{I}(m_1,m^-_{1I},m^+_{2R}) 
+ \mathscr{I}(m_1,m^-_{1R},m^-_{2R}) +
\mathscr{I}(m_1,m^-_{1I},m^-_{2I}) \nonumber \\
&& \qquad\qquad\qquad + \mathscr{I}(m_1,m^+_{1I},m^-_{2R}) +
\mathscr{I}(m_1,m^-_{1R},m^+_{2I}) \biggr]
\nonumber \\
& + &  \frac{\alpha\beta (\gamma_2-\gamma_1)}{128\sqrt{2}\pi^2}m_1 \langle Z
\rangle \biggl[ \mathscr{I}(m_1,m^-_{1R},m^+_{2R})
+\mathscr{I}(m_1,m^+_{1R},m^-_{2R}) +
\mathscr{I}(m_1,m^+_{1I},m^+_{2R}) \nonumber \\
&& \qquad\qquad\qquad + \mathscr{I}(m_1,m^-_{1I},m^-_{2R}) 
+  \mathscr{I}(m_1,m^+_{1R},m^+_{2I})
+ \mathscr{I}(m_1,m^-_{1R},m^-_{2I}) \nonumber \\
&&\qquad\qquad\qquad 
+\mathscr{I}(m_1,m^-_{1I},m^+_{2I})
+ \mathscr{I}(m_1,m^+_{1I},m^-_{2I}) \biggr] \nonumber \\
&-& \frac{\alpha\beta
  m_1m_2}{32\sqrt{2}\pi^2}\biggl[\mathscr{I}(m_1,m_2,m^+_{1R})
+\mathscr{I}(m_1,m_2,m^+_{1I})+\mathscr{I}(m_1,m_2,m^-_{1R})
+\mathscr{I}(m_1,m_2,m^-_{1I})\biggr] \nonumber \\[8pt]
&&\phantom{line}
\eeqa%
where the triangle integral $\mathscr{I}(a,b,c)$ is defined in \eq{iabc}.

In the limit of $\gamma_1=\gamma_2\equiv\gamma$, 
the above results simplify significantly, and the messenger masses
are given by:
\beqa
m_1^2&=& m_2^2=\gamma^2 \langle Z \rangle^2\,,\\
(m^+_{1I})^2&=&(m^-_{1R})^2=(m^+_{2I})^2=(m^-_{2R})^2=\gamma^2 \langle
Z \rangle^2 -\gamma F_Z\,,\\
(m^-_{1I})^2&=&(m^+_{1R})^2=(m^+_{2R})^2=(m^-_{2I})^2=\gamma^2 \langle
Z \rangle^2 + \gamma F_Z\,,
\eeqa% 
in which case \eq{k3eff} simplifies to:
\begin{Eqnarray} \label{k3equalmass}
\frac{k_3}{g} & = & \frac{ \sqrt{2}\alpha \beta \gamma^2
\langle Z \rangle^2}{32\pi^2}\biggl[\mathscr{I}
\left((\gamma^2 \langle Z \rangle^2
+\gamma F_Z)^{1/2},\gamma \langle Z \rangle\right) 
+ \mathscr{I}\left((\gamma^2 \langle Z
\rangle^2-\gamma F_Z)^{1/2},\gamma \langle Z \rangle\right) \nonumber \\
&& - \mathscr{I}\left(\gamma \langle Z \rangle,
(\gamma^2 \langle Z \rangle^2+\gamma F_Z)^{1/2}\right) 
-\mathscr{I}\left(\gamma \langle Z \rangle,
(\gamma^2 \langle Z \rangle^2- \gamma F_Z)^{1/2}\right)\biggr]\,,
\end{Eqnarray}%
where
\beq
\mathscr{I}(a,b) \equiv\mathscr{I}(a,a,b)=\mathscr{I}(b,a,a)=\frac{a^2(a^2-b^2)
+a^2b^2\ln(b^2/a^2)}{a^2(a^2-b^2)^2}\,.
\eeq
An explicit evaluation of \eq{k3equalmass} yields
\beq \label{k3gamma}
\frac{k_3}{g}= \frac{\sqrt{2}\alpha \beta}{32\pi^2}f(x)\,,\qquad \quad 
x\equiv\frac{F_Z}{\gamma \langle Z \rangle^2}\,,
\eeq
where 
\beq
f(x)\equiv\frac{(x-2)\ln(1-x)-(x+2)\ln(1+x)}{x^2}\,.
\eeq
The small $x$ expansion of $f(x)$ gives:
\beq
f(x) = \frac{x^2}{3} +\frac{4x^4}{15} + \mathcal{O}(x^6)\,, 
\eeq
which confirms the behavior of $k_3$ for $x\ll 1$ given in \eq{fzlarge}.
Note that $f(x)\to\infty$ as $x\to 1$, which reflects the fact that
one of the messenger masses is approaching zero.  Thus, we cannot take
$x$ as large as 1.  

We shall choose $x$ such that the lightest messenger mass lies above 1~TeV.
With this bound, $x$ can assume values quite close to 1.  As an example,
consider the case of $\gamma=1$ and 
$F_z/\langle Z \rangle\sim20$~TeV in order that squark and
gaugino masses lie in the appropriate mass range.  If $x$ is close to 1,
then $F_Z \sim \langle Z\rangle^2$, in which case, 
$\langle Z \rangle = 20$~TeV.  If 
$x \simeq 0.98$, then for $\gamma=1$ 
the lightest messenger has a mass of 2.8~TeV.  This is as large an
$x$ value that one could sensibly allow.  At this
particular point in parameter space $f(0.98)=2.0$. This yields a
value for the effective contribution to $k_3/g$ of 
\beq \frac{k_3}{g}= 2.0
\frac{\sqrt{2}\alpha \beta}{32\pi^2}\,. 
\eeq 
This is roughly a maximum possible value for these threshold effects.
Using the formulae given above, we can compute $k_3/g$ for 
a variety of sample points in the parameter space.
In Table~\ref{sample}, we provide four representative points.
The conclusion we draw from this small sample is that with
Yukawa couplings of messengers to Higgs boson of $\mathcal{O}(1)$, 
and the lightest messenger mass above 2~TeV, it is typical
to find values of $k_3/g\sim (0.1$---$1.4)/(16\pi^2)$.  Such corrections
are of one-loop order in size---small but not too small.  
%In \sect{measure}, we shall identify observables that are sensitive
%to $k_3$ as well as being $\tan\beta$-enhanced.

\begin{table}[t!]
\centering
\begin{tabular}{|cclcc|} \hline
\qquad $\gamma_1$\qquad\qquad & \qquad $\gamma_2$\qquad \qquad &\qquad 
$F_Z$\qquad & \qquad\qquad $M_-$\qquad \qquad\qquad
&$16\pi^2 k_3/g$ \qquad \\ \hline
1 & 1 & (19.8~TeV)$^2$& 2.8~TeV & 1.44 \\
0.9 & 1 & (18.8~TeV)$^2$ & 2.4~TeV&   1.38 \\
1 & 1 & (16.8~TeV)$^2$ & 10.9~TeV & 0.19 \\
0.75 & 1 & (14~TeV)$^2$  & 8.8~TeV &  0.15 \\ \hline
\end{tabular}
\caption{\label{sample} Sample points in the messenger parameter space.
We have fixed $\langle Z\rangle=20$~TeV and $\alpha=\beta=1$.
The mass of the lightest messenger state is denoted by $M_-$.}
\end{table}

We have established that in a theory with a low SUSY-breaking
scale in a simple gauge mediated scenario in which messenger
fields couple to the Higgs bosons of the MSSM,
there are dimension-four SUSY-breaking wrong-Higgs gaugino operators
operators [cf.~\eq{k1}-\eq{k4}] generated as threshold corrections at one
loop order. Therefore, one should include these effects in the 
chargino/neutralino sector of the effective
Lagrangian of the MSSM below the fundamental SUSY-breaking
scale. The usual supersymmetric relations between the parameters of
the chargino/neutralino sector and the gauge sector of the MSSM will
then be modified.   
In \sect{measure}, we shall demonstrate
that such effects can be 
considerably enhanced if the parameter $\tan{\beta}$ is large.

\subsection{Renormalization group improvement}

The effective Lagrangian describing the gaugino sector for the MSSM
just below the scale of fundamental SUSY-breaking is given by:
\begin{Eqnarray}\label{gMSSMeff}
\mathscr{L}^{eff}_{gaugino} & = &
\frac{ig_u}{\sqrt{2}}\lambda^a \tau^a_{ij}\psi^j_{H_u}H_u^{*i} +
\frac{ig_d}{\sqrt{2}}\lambda^a \tau^a_{ij}\psi_{H_d}^jH_d^{*i}
+  \frac{ig'_u}{\sqrt{2}}\lambda'\psi_{H_u}^iH_u^{*i}
- \frac{ig'_d}{\sqrt{2}}\lambda'\psi_{H_d}^iH_d^{*i} \nonumber \\
&  - &  M\lambda^a\lambda^a - M'\lambda'\lambda'
-  \mu \epsilon_{ij}\psi^i_{H_u}\psi^j_{H_d}
+ik_1\lambda^a\tau^a_{ij}\psi^j_{H_u}\epsilon_{ki}H_d^k \nonumber \\
 & + &   ik_2 \lambda' \psi_{H_u}^k\epsilon_{ki} H_d^i
+ik_3\lambda^a\tau^a_{ij}\psi_{H_d}^j\epsilon_{ki}H_u^k
+ik_4\lambda'\psi^i_{H_d}\epsilon_{ki}H_u^k + {\rm h.c.}\,,
\end{Eqnarray}%
where we have added the dimension-four wrong-Higgs gaugino operators
given by \eqst{k1}{k4} to the tree-level gaugino Lagrangian [\eq{gMSSM}].
The effective Lagrangian displayed in \eq{gMSSMeff} is 
defined at the threshold scale of the messengers. 
We then use the renormalization group (RG) to run down to the
electroweak scale.
In general the
messengers decouple in two stages: once at the scale $[\langle Z
\rangle^2 + F]^{1/2}$ and once at the scale 
$[\langle Z \rangle^2 - F]^{1/2}$.  For
simplicity, we will estimate the effects of the RG by decoupling the
messengers at the scale $\mu_M = \langle Z \rangle$.  However,
in the limit where the lightest messenger state is extremely light,
two stages of decoupling must be used. Our goal here is to estimate
the effects of the RG analysis on the results from the threshold loops
obtained in the previous section.

We begin with $\mathscr{L}^{MSSM}_{eff}(\mu_M)$ as given in
\eq{gMSSMeff}.  The parameters that appear in this Lagrangian
are effective parameters.  For example,  
$g_u=g + \delta g_u$, $g_d=g + \delta g_d$, $g'_u = g' +
\delta g'_u$, and $g'_d = g' + \delta g'_d$, where the $\delta g's$
include threshold and renormalization group effects from SUSY breaking
below the fundamental SUSY-breaking scale. 
For $M'$, $M$, and $\mu$ we simply
absorb renormalization and threshold corrections into these
coefficients.  In the previous section, we presented an explicit calculation
for $k_3/g$.  The other coefficients
$k_1/g$, $k_2/g'$, and $k_4/g'$ are also generated
with the same order of magnitude.\footnote{We do not present an explicit
calculation of $k_1$, $k_2$ and $k_4$ here.
Instead, we narrow our focus to the chargino sector and in
particular the chargino mass matrix. The coefficients $k_2$ and $k_4$
only affect the neutralino mass matrix so for the subsequent analysis we do not
need the coefficients of these operators.  In \sect{measure}, we demonstrate
that the effects of the wrong-Higgs gaugino
operator proportional to $k_1$ are suppressed at large $\tan{\beta}$, and
can likewise be neglected.}

Because SUSY is broken by dimension-four hard breaking
operators, the theory below $\mu_M$ is non-supersymmetric and the RG for all
couplings must be evolved 
independently~\cite{Machacek:1983tz,Machacek:1983fi,Machacek:1984zw}. 
For all supersymmetric
tree-level couplings, it is a very good approximation to neglect
the presence of the new couplings $k_i$, as
these new couplings are one loop-suppressed.  Moreover,
the wrong-Higgs gaugino operators break the
R-symmetry by 2 units of R-charge (with the standard R-charge
assignments to the regular MSSM superfields). Therefore, the
supersymmetric RG equations for the MSSM couplings are always modified
by terms proportional to the square of $k_i$ (corresponding
diagrammatically to a change of R-charge by $\pm 2$ at the two
wrong-Higgs interaction vertices, respectively).  Therefore the resulting
contribution to the effective supersymmetric coupling is always
suppressed by a factor of 
$\mathcal{O}(1/(16\pi^2)^3)$, which is negligible.  The $k_i$
also evolve according to the RG, and the R-charge analysis implies
that they satisfy RG equations that are linear 
and trilinear in the $k_i$.  The RG equations for the
couplings $k_i$ (neglecting the deviation of
the couplings $g_u$, $g_d$ [and $g_u^\prime$, $g_d^\prime$]
from their supersymmetric values $g$ [and $g'$], respectively)
are given by:
\beqa
\!\!\!\!\!\!\!\!\!\!\!
16\pi^2\frac{dk_1}{dt}&=&\half k_1(6h_t^2 + 6h_b^2
+ 2h_{l}^2+11k_1^2+3k_2^2+2k_3^2)+(g^2-g'^2)k_3 -g'gk_2\,,
\\
\!\!\!\!\!\!\!\!\!\!\!
16\pi^2\frac{dk_2}{dt}&=&\half k_2(6h_t^2+6h_b^2+2h_l^2
+ 2k_4^2 + 9k_1^2+5k_2^2)+3g^2k_4-3g'gk_1+g'^2(k_4+12k_2)\,,
\\
\!\!\!\!\!\!\!\!\!\!\!
16\pi^2\frac{dk_3}{dt}&=&\half k_3(6h_t^2 + 6h_b^2
+ 2h_{l}^2+11k_3^2+3k_4^2+2k_1^2)+(g^2-g'^2)k_1 +g'gk_4\,,
\label{k3rge}\\
\!\!\!\!\!\!\!\!\!\!\!
16\pi^2\frac{dk_4}{dt}&=&\half k_4(6h_t^2+6h_b^2+2h_l^2
+ 2k_2^2 + 9k_3^2+5k_4^2)+3g^2k_2+3g'gk_3+g'^2(k_2+12k_4)\,.
\eeqa%

If one keeps only the
largest terms in the RG equation for $k_3$, then \eq{k3rge} reduces to:
\beq 16\pi^2\frac{dk_3}{dt}=
k_3(3h_t^2 + 3h_b^2)\,.
\eeq
As an example, for $\tan{\beta}=50$, we obtain $h_t=0.95$ and
$h_b=1.16$.  This provides a first estimate of the RG correction to $k_3$
\beq \label{RGrel} 
0.86\,k_3(\mu_M=20~{\rm TeV})=  k_3(\mu=500~{\rm GeV}). 
\eeq
That is, RG-evolution has reduced the size of $k_3$ (obtained in
\sect{largesquark}) by roughly $14\%$.
More generally, we expect
modifications of the threshold values of the $k_i$ to be of order $10\%$ by
RG running in the parameter regime of interest.

\section{Effects of wrong-Higgs chargino operators on the chargino mass matrix}
\label{measure}

\subsection{Dimension-four hard SUSY-breaking corrections to the
  chargino mass matrix} 
\label{tanbenh}

After the neutral Higgs bosons acquire their vacuum expectation values,
$\langle H_u^0\rangle=v_u/\sqrt{2}$ and  $\langle H_d^0\rangle=v_d/\sqrt{2}$,
the quadratic terms of the effective gaugino Lagrangian [\eq{gMSSMeff}]
are given by:
\beqa
\mathscr{L}_{\rm mass} & =& 
\frac{ig_u v_u}{2}\lambda^a \tau^a_{2j}\psi^j_{H_u} +
\frac{ig_d v_d}{2}\lambda^a \tau^a_{1j}\psi_{H_d}^j
+  \frac{ig'_u v_u}{2}\lambda'\psi_{H_u}^2
- \frac{ig'_d v_d}{2}\lambda'\psi_{H_d}^1 
 -   M\lambda^a\lambda^a - M'\lambda'\lambda' \nonumber \\
&-&  \mu \epsilon_{ij}\psi^i_{H_u}\psi^j_{H_d}
+\frac{ik_1 v_d}{\sqrt{2}}\lambda^a\tau^a_{2j}\psi^j_{H_u} 
  -    \frac{ik_2 v_d}{\sqrt{2}} \lambda' \psi_{H_u}^2
- \frac{ik_3v_u}{\sqrt{2}}\lambda^a\tau^a_{1j}\psi_{H_d}^j
- \frac{ik_4 v_u}{\sqrt{2}}\lambda'\psi^1_{H_d} + {\rm h.c.}\nonumber \\
\phantom{line}
\eeqa%
Isolating the terms that contribute to the chargino matrix, we
introduce
\beq
\psi^+_i = \left( \begin{array}{c} -i\lambda^+ \\
\psi_{H_u}^1 \end{array} \right)\,, \qquad\qquad \psi^-_i
= \left( \begin{array}{c} -i \lambda^- \\ \psi_{H_d}^2 \end{array} \right)\,,
\eeq
where $\lambda^\pm=\frac{1}{\sqrt{2}}(\lambda^1\mp i\lambda^2)$.
Then, the chargino mass terms are given by:
\beq
\mathscr{L}_{\rm mass} = -\frac{1}{2} \left( \begin{array}{cc} \psi^+
& \psi^-  \end{array} \right) \left( \begin{array}{cc} 0 & (X^{\rm eff})^T \\
X^{\rm eff} & 0  \end{array} \right) \left( \begin{array}{c} \psi^+ \\ \psi^-
\end{array} \right) + {\rm h.c.}\,,
\eeq
where
\beq \label{X}
 X^{\rm eff}  \equiv \left( \begin{array}{cc}  
X^{\rm eff}_{11}\quad  & X^{\rm eff}_{12} \\
X^{\rm eff}_{21} \quad & X^{\rm eff}_{22} \end{array} \right)=
\begin{pmatrix} M  & (g+\delta g_u)\displaystyle\frac{v_u}{\sqrt{2}}
\left( 1-\displaystyle
\frac{\sqrt{2}k_1\cot\beta}{g+\delta g_u}\right) \\
(g+\delta g_d)\displaystyle\frac{v_d}{\sqrt{2}}\left( 1+\displaystyle
\frac{\sqrt{2}k_3\tan\beta}{g+\delta g_d}
\right)  & \mu\end{pmatrix}
\eeq
with $v_u\equiv v\sin\beta$ and $v_d\equiv v\cos\beta$.

In the limit of large $\tan{\beta}$, the correction to the
supersymmetric relation, $X_{21} = gv\cos{\beta}/\sqrt{2}$, is
significant. Including effects from the improved renormalization group
running of the parameters of Table~\ref{sample}, this correction can
be as large as $7\%$---$56\%$ for $\tan{\beta}=50$ as $F_Z$ varies
between $14$--19~TeV.  In this estimate we have neglected the effects
of $\delta g_d$ as these are one-loop effects with no
$\tan\beta$-enhancements.

In~\cite{Feng:1995zd,Choi:1998ut,Choi:1998ei,Choi:2000ta}, it was
shown in detail how to extract the parameters of the chargino sector
from polarized $e^+e^-$ experiments.  By employing these techniques,
it should be possible to detect deviations from the standard
MSSM expectations.  We have seen above that the effect of
the wrong-Higgs chargino operators is to generate a potentially
significant $\tan\beta$-enhanced correction to the supersymmetric
value of $X_{21}$.  Hence, we focus on the perturbation of the chargino
mass matrix due to a shift in the value of $X_{21}$.

\subsection{A perturbative analysis of 
the contribution of the wrong-Higgs gaugino couplings to 
the chargino mass matrix}

Given the effective chargino mass matrix of \eq{X} [henceforth denoted
as $X$], we can compute the chargino eigenvalues and corresponding
diagonalization matrices.  Any complex matrix possesses 
a singular value decomposition~\cite{horn} of the form:
\beq \label{svd}
U^*XV^{-1} = M_D\equiv {\rm diag}(m_{\chi_1^+},m_{\chi_2^+})\,,
\eeq
for some suitably chosen unitary matrices $U$ and $V$, where the
elements of the diagonal matrix $M_D$ are real and non-negative.
Note that \eq{svd} implies that:
\beq \label{svd2}
VX^{\dagger}XV^{-1}=U^*XX^{\dagger}U^T = M^2_D\,.
\eeq
Thus, the chargino masses are determined by solving the eigenvalue
problem for either $X^\dagger X$ (or equivalently, for $XX^\dagger$).
Moreover, to compute the unitary matrices $U$ and $V$, one can
first determine the matrix $U$ by diagonalizing $XX^\dagger$ and then
compute $V$ from \eq{svd} [or equivalently, one can first determine
the matrix $V$ by diagonalizing $X^\dagger X$ and then compute $U$
from \eq{svd}].  In the former procedure, $U$ is unique up to 
multiplication on the right by a diagonal matrix of phases (assuming
that the elements of $M_D$ are non-degenerate).  We shall use this
phase freedom to reduce the number of parameters of the unitary matrix
$U$ from four to two; that is, we can parameterize $U$ as 
follows~\cite{Choi:1998ut,Choi:1998ei,Choi:2000ta}:
\beq
U = \left( \begin{tabular}{cc} $\cos{\theta_L}$&
$e^{-i\beta_L}\sin{\theta_L}$ \\ $-e^{i\beta_L}\sin{\theta_L}$&
$\cos{\theta_L}$ \end{tabular} \right)\,.
\eeq
Once $U$ has been fixed, then $V$ is uniquely determined by \eq{svd}.
The unitary matrix $V$ depends on four parameters, which we
parameterize as:
\beq
V =  \left( \begin{tabular}{cc} $e^{i\zeta_1}$& $0$ \\ $0$&
$e^{i\zeta_2}$ \end{tabular} \right)  \left( \begin{tabular}{cc}
$\cos{\theta_R}$& $e^{-i\beta_R}\sin{\theta_R}$ \\
$-e^{i\beta_R}\sin{\theta_R}$& $\cos{\theta_R}$ \end{tabular}
\right)\,.
\eeq

In the MSSM (with the hard-breaking SUSY contributions set to zero),
the only non-trivial phase is the relative phase between $M$ and
$\mu$.  One can always absorb the phase of $M$ by rephasing the
SU(2) gaugino field, in which case the only remaining potentially
complex parameter is $\mu\equiv |\mu|e^{i\Phi}$.  Thus, without loss
of generality, we take $M$ real and positive.  
New phases can also enter due to the complexity of
the parameters $\gamma_1$, $\gamma_2$,
$\alpha$ and $\beta$ that parameterize the messenger
superpotential [cf.~\eq{spotential}].  Consequently, in the chargino mass
matrix, two new independent phases can appear in $X_{12}$ and
$X_{21}$.   For simplicity, we assume in what follows that these
phases are either absent or negligible.  We shall address
the implication of non-negligible CP-violating phases in $X_{12}$ and
$X_{21}$ in a future publication.

For $M>0$ and $X_{12}$ and $X_{21}$ real,
the chargino squared-masses and mixing angles $\theta_L$ and
$\theta_R$ are easily obtained:
\beqa
m^2_{\chi^{\pm}_{1,2}}&=& \half \left( M^2 + |\mu|^2
+X_{12}^2+ X_{21}^2 \mp \Delta \right)\,,\label{chimass}\\[6pt]
\cos{2\theta_{R,L}} &=&\frac{|\mu|^2- M^2 \pm (X_{12}^2 -X_{21}^2)}
{\Delta}\,,\label{chimix}
\eeqa%
where the quantity $\Delta$ is defined by:
\beq \label{Deltadef}
\Delta\equiv\left[(M^2- |\mu|^2 -X_{12}^2 + X_{21}^2)^2 +4(M^2X_{12}^2
+|\mu|^2X_{21}^2 +2M|\mu| X_{12}X_{21}\cos{\Phi})\right]^{1/2}\,.
\eeq
The four phase angles $\beta_L$, $\beta_R$, $\zeta_1$ and
$\zeta_2$ are given by:
\beqa
\tan\beta_L&=&\frac{- X_{12}|\mu|
 \sin\Phi}{X_{21}M+X_{12}|\mu|\cos\Phi}\,,\label{angle1}\\[7pt]
\tan\beta_R&=&\frac{ X_{21}|\mu|
 \sin\Phi}{X_{12}M+X_{21}|\mu|\cos\Phi}\,,\label{angle2}\\[7pt]
\tan\zeta_1&=&\frac{X_{12}X_{21}|\mu|\sin\Phi}{X_{12}X_{21}|\mu|\cos\Phi
+M(m^2_{\chi^{\pm}_{1}}-|\mu|^2)}\,,\label{angle3}\\[7 pt]
\tan\zeta_2 &=& \frac{-(m^2_{\chi^{\pm}_{2}}-M^2)|\mu|\sin\Phi}
{(m^2_{\chi^{\pm}_{2}}-M^2)|\mu|\cos\Phi+X_{12}X_{21}M}\,.\label{angle4}
\eeqa%
\Eqst{chimass}{angle4} are a simple extension of the MSSM results obtained
in~\cite{Choi:1998ut}.
It is convenient to define the following quantities:
\beq
C_{RL}^+ \equiv -(\cos{2\theta_R} + \cos{2\theta_L})\,,\qquad\quad
C_{RL}^- \equiv \cos{2\theta_R} - \cos{2\theta_L}\,.
\eeq
Then, \eqs{chimass}{chimix} are equivalent to the following four relations: 
\begin{Eqnarray} 
C^+_{RL} (m^2_{\chi^{\pm}_2} - m^2_{\chi^{\pm}_1}) &=& 2(M^2 -\mu^2)\,,
\label{matrixrel1} \\
C^-_{RL} (m^2_{\chi^{\pm}_2} - m^2_{\chi^{\pm}_1})
&=& 2(X_{12}^2-X_{21}^2)\,,
\label{matrixrel2} \\
m^2_{\chi^{\pm}_2} + m^2_{\chi^{\pm}_1} &=& M^2 +\mu^2 + X_{12}^2 + X_{21}^2\,,
\label{matrixrel3} \\
\Delta &=& m^2_{\chi^{\pm}_2} - m^2_{\chi^{\pm}_1}\,.\label{matrixrel4}
\end{Eqnarray}%

In the absence of dimension-four hard-SUSY-breaking operators, the
tree-level values of the off diagonal elements of $X$ are
given by $X_{12}=\sqrt{2}m_W\sin{\beta}$
and $X_{21}=\sqrt{2}m_W\cos{\beta}$. Including corrections due to
the hard SUSY-breaking operators, the chargino matrix is modified by
small corrections (of one-loop order), which can be treated perturbatively.
That is, we write:
\beqa 
X_{12}&=&\sqrt{2}m_W\sin{\beta}(1+\delta_{12})\,,\\
X_{21}&=&\sqrt{2}m_W\cos{\beta}(1+\delta_{21})\,,
\eeqa%
where $\delta_{12}$
and $\delta_{21}$ are small, and we work to first order in
these small quantities.  Our ultimate goal to is express $\delta_{12}$
and $\delta_{21}$ in terms of the chargino masses
$m_{\chi^{\pm}_{1,2}}$, the ratio of Higgs vacuum expectation values,
$\tan\beta$, the mixing angles $\theta_L$ and $\theta_R$,
and the phase of $\mu$ (denoted above by $\Phi$).  In principle, these
quantities can be determined by precision measurements of the chargino
system as described in~\cite{Choi:2000ta}.

We first rewrite \eq{matrixrel2} as:
\beq \label{firsteq}
s_\beta^2\delta_{12}-c_\beta^2\delta_{21}=\half c_{2\beta}+
\frac{C^-_{RL}(m^2_{\chi^{\pm}_{2}}-m^2_{\chi^{\pm}_{1}})}{8m_W^2}\,,
\eeq
where $s_\beta\equiv\sin\beta$, $c_\beta\equiv\cos\beta$, etc.
We next use \eqs{matrixrel1}{matrixrel3} to solve for $M$ and $|\mu|$.
Inserting these results into \eq{Deltadef} yields a second
linear equation for $\delta_{12}$ and $\delta_{21}$ (after dropping
higher order terms in the $\delta$'s) of the
following form:
\beq \label{deltarel} 
g\delta_{21} + h\delta_{12} = 2f^{1/2}(\Delta -f^{1/2})\,,
\eeq 
where
\begin{Eqnarray}
f  &=&   (\half C_{RL}^+\Delta+ 2m_W^2c_{2\beta})^2
+ 4m_W^2(m^2_{\chi^{\pm}_{2}} + m^2_{\chi^{\pm}_{1}}
-2m_W^2)-2m_W^2C_{RL}^+\Delta c_{2\beta} 
 +  4 m_W^2\Gamma s_{2\beta}\cos{\Phi}\,,\nonumber \\ 
&& \phantom{line} \\
g & = &   2m_W^2c_{\beta}^2 \biggl[ 4(m^2_{\chi^{\pm}_{2}}
+ m^2_{\chi^{\pm}_{1}})
+ 4m_W^2c_{2\beta}-16m_W^2-C_{RL}^+\Delta 
+ 4\Gamma\tan{\beta}\cos{\Phi} \nonumber \\
&& \hspace{1in}
-\frac{8m_W^2}{\Gamma}(m^2_{\chi^{\pm}_{2}} + m^2_{\chi^{\pm}_{1}}
-2m_W^2)s_{2\beta}\cos{\Phi}\biggr]\,, \\[8pt]
h  &=&  2m_W^2s_{\beta}^2 \biggl[ 4(m^2_{\chi^{\pm}_{2}}
+ m^2_{\chi^{\pm}_{1}}) - 4m_W^2c_{2\beta}
-16m_W^2+C_{RL}^+\Delta +4\Gamma\tan{\beta}\cos{\Phi} \nonumber \\
&& \hspace{1in} -\frac{8m_W^2}{\Gamma}(m^2_{\chi^{\pm}_{2}}
+ m^2_{\chi^{\pm}_{1}} -2m_W^2)s_{2\beta}\cos{\Phi}\biggr]\,,
\end{Eqnarray}%
with
\beq
\Gamma\equiv \left[(m^2_{\chi^{\pm}_{1}} + m^2_{\chi^{\pm}_{2}} -2m_W^2)^2
-\quarter (C_{RL}^+\Delta)^2 \right]^{1/2}\,.
\eeq

\Eqs{firsteq}{deltarel} provide two equations for the unknowns
$\delta_{12}$ and $\delta_{21}$.  Solving for $\delta_{21}$, we find:
\beq \label{susyrel}
\delta_{21}=\frac{2s_\beta^2 f^{1/2}(\Delta-f^{1/2})
-\half h\left[c_{2\beta}
+\frac{\displaystyle C^-_{RL}(m^2_{\chi^{\pm}_{2}}-m^2_{\chi^{\pm}_{1}})}
{\displaystyle 4m_W^2}\right]}{hc_\beta^2+gs_\beta^2}\,.
\eeq
As a check, consider the supersymmetric limit where
$X_{12}/X_{21}=\tan\beta$ and $X_{12}X_{21}=m_W^2 s_{2\beta}$.
In this limit, $f=\Delta^2$ and 
$C^-_{RL}(m^2_{\chi^{\pm}_{2}}-m^2_{\chi^{\pm}_{1}})=-4m_W^2
c_{2\beta}$.  Hence, $\delta_{21}=0$ in the limit of exact 
supersymmetry as expected.

We have achieved our goal of expressing $\delta_{21}$ in terms of 
the chargino masses, $\tan\beta$, the mixing angles $\theta_L$ 
and $\theta_R$, and $\Phi=\arg\mu$.  In~\cite{Choi:2000ta}, it is
shown how to extract the values of the chargino masses and the mixing angles
$\theta_L$ and $\theta_R$ and $\Phi$ from precision chargino data at
the International Linear Collider (ILC)
in a model-independent way, using measurements of the total production
cross-sections for $e^+e^-\to\widetilde\chi_i^\pm\widetilde\chi_j^\mp$
and asymmetries with polarized beams. (A similar proposal for
measuring the chargino
masses and mixing angles in a CP-conserving scenario was put forward
in \cite{Feng:1995zd}.)
If $\tan\beta$ and $\Phi=\arg\mu$ are
known independently, then \eq{susyrel} provides a prediction for
$\delta_{21}$. 
For example, $\tan\beta$ can be determined from
precision Higgs measurements (if the heavy Higgs states are observed).
An independent determination of $\Phi$ is more problematical.
Within the context of the MSSM chargino sector, 
it is shown in~\cite{Choi:2000ta} that one can also extract values for
$\tan\beta$ and $\Phi$ from the precision chargino data.  But, this
determination relies on the standard MSSM chargino mass matrix 
where $\delta_{12}=\delta_{21}=0$.  This procedure must be generalized
if the $\delta$'s are nonzero.
In principle, it should be possible
to solve for all the unknown quantities if the appropriate
linear combinations of the phases $\beta_L$, $\beta_R$, $\zeta_1$ and
$\zeta_2$ can be determined experimentally.\footnote{A similar
analysis of the neutralino sector can provide important cross checks
of the SUSY parameter determination.  We will address these points in
more detail in a future publication.}

Thus, a measurement of the effective chargino mass matrix in this way
can signal an effect of SUSY-breaking physics beyond the MSSM.  These
conclusions depend on the assertion that the tree-level effects of the
dimension-four hard-SUSY-breaking operators dominate the more generic
loop corrections of pole masses and interactions that cannot be
described by terms in a local effective Lagrangian.  We discuss the
validity of this assumption in the next section.

\section{Local versus non-local effects}
\label{nonlocal}

In this paper, we
we have analyzed the local effective Lagrangian generated by
including a low-scale messenger sector that couples via Yukawa interactions
to the Higgs doublets.  As a consequence, dimension-four wrong-Higgs
gaugino interactions are generated with a strength proportional to
the product of messenger--Higgs Yukawa couplings,
$\alpha \beta$. These couplings then enter the chargino mass matrix,
thereby perturbing the standard MSSM relations satisfied by chargino mass
matrix elements.  However, the chargino masses and mixing angles are
also modified at one-loop due to momentum-dependent
radiative corrections in which the
MSSM fields propagate in the loop.  Such effects
have been thoroughly investigated in
various regions of the MSSM parameter 
space~\cite{Fritzsche:2002bi,Eberl:2001eu,Oller:2003ge}. 
These ``non-local'' effects can
compete with the local effects of the hard-SUSY-breaking operators
in certain regimes of parameter space, and have not been included in
\eq{susyrel}.  Here, we shall argue that in our GMSB
scenario, it is possible for the local effects to dominate if the 
messenger--Higgs Yukawa couplings are larger than the 
electroweak gauge couplings.

First, consider the one-loop corrections to the 
off-diagonal elements of the chargino mass matrix
arising from squark exchange.  Examples of such corrections
are depicted in \fig{failhiggs}, after the Higgs field acquires a
vacuum expectation value.  In \sect{within}, we demonstrated that
these corrections decouple at large squark masses\footnote{As
discussed in \sect{largesquark}, in the limit where squarks decouple,
the one-loop wrong-Higgs gaugino interaction actually arises from
a local dimension-six operator.  The corresponding local wrong-Higgs 
gaugino operators are generated when the neutral Higgs bosons take on
their vacuum expectation values.}  
and are additionally
suppressed by a factor of the bottom-quark Yukawa coupling.  Indeed,
in GMSB scenarios, the squark masses are expected to be rather large,
as one generically expects mass relations of the form
\beq
\frac{m_{\tilde{e}^{\pm}_R}}{m_{\tilde q}} \sim \frac{g'^2}{g_3^2}\,.
\eeq
Using the current lower bound on the selectron mass of
$m_{\tilde e_R^{\pm}} > 73$~GeV~\cite{pdg}, 
it follows that squarks should be  
quite heavy, $m_{\tilde q} > 800$~GeV.  
As a result, we do not expect the
squark-exchange contributions to be significant.

Next, we consider the effects of virtual slepton, chargino, and
neutralino exchange at one-loop that can contribute
competing non-local effects to the
wrong-Higgs operators of interest.  Here we note that any loop correction to
chargino/neutralino masses and interactions with charginos/neutralinos
or sleptons/leptons propagating on the internal lines will enter with
at least two factors of the electroweak couplings $g$ and $g^{\,\prime}$, for
the chargino/neutralino contributions or the lepton Yukawa couplings
for the lepton/slepton contributions. As a result, the only
important non-local effects are $\sim g^{\,\prime\, 2}$ and $\sim g^2$
competing against effects $\sim \alpha
\beta$ from the messenger sector. 
As long as $\alpha \beta > g^2$, $g^{\,\prime\, 2}$, the
messenger effects will always be parametrically larger than the
non-local corrections.  Thus, with the assumption that $\alpha \beta >
g^2$, $g^{\,\prime\, 2}$, 
a measurement of a significant deviation of $\delta_{21}$
from zero means that the measured deviation is coming from effects
beyond the MSSM.   
The gauge-mediated model with messenger--Higgs
Yukawa couplings provides a plausible scenario in which non-negligible
effects in the chargino sector due to the messenger sector are possible.

\section{Conclusions and Outlook}
\label{conclude}

In models of low-energy supersymmetry, there is often a hierarchy of
scales that governs the structure of the model.  At scales above
2.5~TeV, messenger fields can provide an avenue for the communication
of the fundamental SUSY-breaking from the hidden sector 
to the visible sector of the MSSM fields.  The scale of the
superpartner masses of the MSSM is roughly determined by
the scale of low-energy SUSY-breaking, which we take to be
$\msusy\sim\mathcal{O}(1$~TeV).  Finally, the electroweak
symmetry-breaking scale $v\sim 246$~GeV provides the masses for the
electroweak gauge bosons and one or more of the Higgs bosons.
At each of the two higher scales, one can integrate out the heavy
states to obtain an effective low-energy Lagrangian, valid at the
electroweak scale.  Some of the physics of SUSY-breaking is then
encoded in dimension-four hard supersymmetry breaking operators that
appear in the low-energy effective Lagrangian.

In this paper, we have focused on the so-called wrong-Higgs couplings
of the MSSM.  These are gauge-invariant dimension-four couplings of
the Higgs bosons to other Standard Model and/or MSSM fields that
violate supersymmetry.  If the low-energy effective Lagrangian
describes the two-Higgs-doublet extension of the Standard Model, then
the wrong-Higgs  couplings are dimension-four Higgs-fermion 
Yukawa couplings 
that violate supersymmetry.  These couplings are generated in one-loop
corrections to the Yukawa interactions due to the exchange of heavy
superpartners in the loops.  The effects of the heavy superpartners
do not decouple if all supersymmetry mass parameters are
simultaneously taken large.  The implication of these
wrong-Higgs interactions include some $\tan\beta$-enhanced corrections
to certain tree-level relations that can be phenomenologically 
important.  

If the low-energy Lagrangian includes the charginos and neutralinos of
the MSSM, then the wrong-Higgs couplings are dimension-four
gaugino--higgsino--Higgs boson couplings that violate supersymmetry.
We have demonstrated that such couplings do \textit{not} arise from
one-loop corrections with heavy squarks in the loop.  The latter
effects decouple as the squark mass is taken heavy, and are derivable
from a dimension-six operator with a coefficient that behaves inversely
with the square of the heavy squark mass.  In models of gauge-mediated
supersymmetry breaking with a low messenger scale, the messenger
fields can have direct couplings to the Higgs bosons.  Consequently,
one must also evaluate one-loop corrections to 
gaugino--higgsino--Higgs boson couplings with the messenger fields in
the loop.  Integrating out the messenger fields yields an
effective low-energy Lagrangian with wrong-Higgs gaugino interactions.
The wrong-Higgs gaugino interactions modify the tree-level chargino
and neutralino mass matrix.

In this paper, we have focused on the effect of the wrong-Higgs
gaugino operators on the chargino mass matrix.  The off-diagonal
elements of this mass matrix are modified from their supersymmetric
values.  For one of the two off-diagonal elements, this deviation is
enhanced at large $\tan\beta$, and can range from a few percent to as
much as 56\% for $\tan\beta=50$.  To detect such a deviation in
experimental data, one would need to initiate a program of precision
chargino measurements in order to reconstruct the underlying
parameters that govern the chargino mass matrix.  Such a program would
begin at the LHC, but the required precision would most likely require
chargino production at the ILC.  A strategy for the reconstruction of
the chargino mass matrix at the ILC has been given in
\cite{Choi:1998ut,Choi:1998ei,Choi:2000ta}, in the case where the
wrong-Higgs gaugino couplings are absent.  We have derived a relation
between observable chargino parameters and the coefficient of the
$\tan\beta$-enhanced wrong-Higgs coupling.  Whether future ILC
chargino data can provide statistically significant evidence for the
wrong-Higgs couplings under realistic experimental conditions requires
further study.

In this paper, we have focused on the implications of the wrong-Higgs
gaugino couplings for the chargino sector.  Similar $\tan\beta$-enhanced
effects due to wrong-Higgs gaugino couplings also modify the off-diagonal
elements of the neutralino mass matrix.  The analysis of these effects
and their phenomenological implications are somewhat more complicated
and will be postponed to a future investigation.

Finally, we note that the existence of the wrong-Higgs gaugino
couplings derived in this paper was a consequence of a very specific
Higgs-messenger interaction which need not be generic in the class of
gauge-mediated supersymmetry-breaking models.  It would be interesting
to classify extensions of the MSSM that yield similar conclusions.
Such an extension would have to possess a field that that experiences
SUSY-breaking, is charged under the electroweak gauge group, and
couples to the Higgs bosons.  On the other hand, one can also take a
purely phenomenological point of view.  Having established that the
wrong-Higgs gaugino couplings do arise in some class of models, one
can simply assume their existence and classify all possible
phenomenological consequences of such operators for supersymmetric
events at future colliders.  Ultimately, if experimental evidence for
such wrong-Higgs operators can be confirmed, such a result would have
a profound impact on the search for the fundamental principles that
govern supersymmetry-breaking.

\section*{Acknowledgements}

We are grateful for a number of illuminating conversations with
Michael Dine and Scott Thomas
on the theory and models of gauge-mediated supersymmetry
breaking.  This work was supported in part by the U.S. Department of
Energy, under grant number DE-FG02-04ER41268.  In addition, J.D.M.
acknowledges the generous support of the ARCS Foundation.

\bibliography{hm-1}

\end{document}